\documentclass[aps,pre,twocolumn,floatfix,showpacs]{revtex4}
\usepackage{graphicx}
\usepackage{bm}
\bibstyle{apsrev.bst}

\begin{document}
\title{Physics-based analysis of Affymetrix microarray data}
\author{T. Heim}
\affiliation{Interdisciplinary Research Institute c/o IEMN, Cit\'e
Scientifique BP 60069, F-59652 Villeneuve d'Ascq, France}
\author{L.-C. Tranchevent}
\affiliation{Interdisciplinary Research Institute c/o IEMN, Cit\'e
Scientifique BP 60069, F-59652 Villeneuve d'Ascq, France}
\author{E. Carlon}
\affiliation{Interdisciplinary Research Institute c/o IEMN, Cit\'e
Scientifique BP 60069, F-59652 Villeneuve d'Ascq, France}
\affiliation{Ecole Polytechnique Universitaire de Lille, Cit\'e
Scientifique, F-59655 Villeneuve d'Ascq, France}
\author{G. T. Barkema}
\affiliation{Institute for Theoretical Physics, University of Utrecht,
Leuvenlaan 4, 3584 CE Utrecht}
\date{\today}

\begin{abstract}
We analyze publicly available data on Affymetrix microarrays spike-in
experiments on the human HGU133 chipset in which sequences are added in
solution at known concentrations.  The spike-in set contains sequences
of bacterial, human and artificial origin.  Our analysis is based on a
recently introduced molecular-based model [E. Carlon and T. Heim, Physica
A {\bf 362}, 433 (2006)] which takes into account both probe-target
hybridization and target-target partial hybridization in solution.
The hybridization free energies are obtained from the nearest-neighbor
model with experimentally determined parameters.  The molecular-based
model suggests a rescaling that should result in a ``collapse" of the
data at different concentrations into a single universal curve.  We indeed
find such a collapse, with the same parameters as obtained before for the
older HGU95 chip set.  The quality of the collapse varies according to
the probe set considered.  Artificial sequences, chosen by Affymetrix
to be as different as possible from any other human genome sequence,
generally show a much better collapse and thus a better agreement with
the model than all other sequences. This suggests that the observed
deviations from the predicted collapse are related to the choice of
probes or have a biological origin, rather than being a problem with
the proposed model.  \end{abstract}


\maketitle

\newcommand{\ul}{\underline}
\newcommand{\bc}{\begin{center}}
\newcommand{\ec}{\end{center}}
\newcommand{\be}{\begin{equation}}
\newcommand{\ee}{\end{equation}}
\newcommand{\ba}{\begin{array}}
\newcommand{\ea}{\end{array}}
\newcommand{\beqn}{\begin{eqnarray}}
\newcommand{\eeqn}{\end{eqnarray}}

\section{Introduction}
\label{sec:intro}

DNA microarrays \cite{sche95} allow to measure the gene expression level
of thousands of genes simultaneously. This is a major step forward
compared to traditional methods in molecular biology (as Northern
blots) which are applicable only to a limited set of genes at a time.
The determination of gene expression levels is not the only application
of DNA microarrays, which have been used also for the analysis of
genetic variance between individuals (single nucleotide polymorphisms),
as efficient tools for DNA sequencing, for the study of chromosomal
defects and for the determination of alternative splicing events.

Despite the increasing popularity that microarrays have known in the
recent years there are still some problems with the technology. There
has been, for instance, only a moderate effort in comparing different
microarrays platforms on the same biological system \cite{mars04}. When
this comparison was made, as in a recent study on expression analysis of
stressed-out pancreas cells, it was found that different commercial
platforms produced wildly incompatible data \cite{tan03_sh}.  These
problems call for a better fundamental understanding of the functioning of
the microarrays. Such understanding will help researchers to design better
algorithms for microarray data analysis based on the physical-chemistry
of the underlying hybridization process.

In the past years several experiments were addressing the analysis
of equilibrium and dynamical properties of DNA hybridization to
probes anchored on solid surfaces with different techniques as,
for instance, surface plasmon resonance \cite{pete02} and by quartz
microbalance \cite{okah98_sh}.  At the same time several papers
\cite{vain02,held03,naef03,haga04,halp04,bind05,carl06} have been dedicated
to theoretical aspects of hybridization, mostly discussing the Langmuir
model and variances thereof.

In a previous paper \cite{carl06} we have analyzed a series of publicly
available data of experiments performed on Affymetrix microarrays, using a
simple model of the hybridization process. In these experiments a set of
selected genes are ``spiked-in" at fixed concentrations into a solution
containing other types of RNAs. This set of data has been widely used
as testground for algorithms designed to extract gene expression levels
from the raw data. Affymetrix is one of the major commercial producers of
microarrays. In Affymetrix arrays the surface-bound probes are prepared in
situ by photolitographic techniques.  Although the technique is limited
to rather short oligos (25 nucleotides long) one of the advantages is
that a high density of probe sequences per array can be obtained. In the
latest generation 1,400,000 different probes have been placed in a single
array. The large number of probes compensate for their limited length.
Indeed Affymetrix uses multiple probes per gene, which define a probe set.
Another special feature of Affymetrix chips is that it uses as control a
mismatch (MM) probe sequence, which differs from a perfect-matching (PM)
sequence only at the base at position 13: a nucleotide A is interchanged
with T and a nucleotide C is interchanged with G.

In our previous work \cite{carl06} we focused on the spike-in data set
of the HGU95 human chipset. More recently this has been substituted
by the HGU133 chipset. Probe sets have been completely redesigned in
the HGU133 chipset; moreover there are only 11 probes per probe set
compared to the 16 probes of the HGU95 array.  In this paper we focus
on the analysis of publicly available spike-in data on the HGU133 chip,
building on our previous work \cite{carl06} on HGU95. This will allow us
to test the robustness of the model introduced in Ref. \cite{carl06}
to a new set of data.  There is another interesting feature of the
spike-in data of the HGU133 chipset: differently from the HGU95 data
where spikes correspond to human genes, the spikes in the HGU133 have
been selected between human, bacterial and ``artificial" sequences.
The latter were selected by Affymetrix to avoid cross-hybridization with
any known human coding sequence.

%%%%%%%%%%%%%%%%%%%%%%%%%%%%%%%%% FIG_01 %%%%%%%%%%%%%%%%%%%%%%%%%%%%%%%%%%%
\begin{figure}[t]
\includegraphics[width=8.5cm]{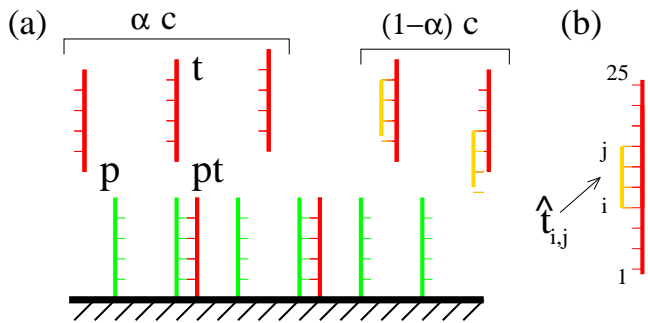}
\caption{(a) The simple model of hybridization in Affymetrix microarrays
used throughout this paper is defined by two basic reactions: 1)
Hybridization between target molecules ({\it t}) to surface anchored
probes ({\it p}) leading to a duplex {\it pt} and 2) The hybridization
between target molecules in solution leading to the partial duplexes
$t {\hat t}_{i,j}$.  In the model, the effect of the hybridization in
solution amounts to a reduction of the original target concentration
$c$ to a value $\alpha c$.  (b) Partial hybridization of a fragment in
solution complementary to the target RNA sequence from base $i$ to base
$j$ ($1 \leq i < j \leq 25$).
}
\label{FIG00}
\end{figure}
%%%%%%%%%%%%%%%%%%%%%%%%%%%%%%%%% FIG_01 %%%%%%%%%%%%%%%%%%%%%%%%%%%%%%%%%%%

\section{A simple model for hybridization in Affymetrix arrays}
\label{sec:model}

In this section we briefly recall the model introduced in
Ref. \cite{carl06}. Two basic processes are considered: 1) Target-probe
hybridization and 2) Target-target hybridization in solution.  According
to the model the fluorescence signal measured from a given probe is:
\be
I = I_0 + \frac{A \alpha c e^{\beta \Delta G}}{1 + \alpha c e^{\beta \Delta G}}
\label{fluorescence}
\ee
where $I_0$ indicates a background level due to non-specific
hybridization, $A$ sets the scale of intensities, $c$ is the target
concentration (a measure of the gene expression level), $\Delta G$
the target/probe hybridization free energy, $\beta = 1/RT$ the
inverse temperature, $R$ the universal gas constant. Here, $\alpha$
models the reduction in the concentration of available targets due to
the target-target hybridization in solution: only a fraction $\alpha
c$ is available for the hybridization with probes as the remaining
$(1-\alpha)c$ form stable duplexes with other partners in solution (see
Fig. \ref{FIG00}(a)).

In the model introduced in Ref. \cite{carl06}, we approximate the
target-target hybridization with the expression
\be
\alpha \approx \frac 1 
{1 + \tilde{c} \exp{\left( \beta' \Delta G_R^{(37)} \right)}}
\label{alpha}
\ee
with $\beta'$ and $\tilde{c}$ fitted parameters and $\Delta G_R^{(37)}
\equiv \Delta G_R (1,25)$ the (sequence dependent) RNA/RNA free energy
for duplex formation in solution at 37 degrees calculated over the whole
25-mer length; in close approximation, the binding free energies at 37 and
45 degrees (the actual experimental temperature) are almost identical,
apart from a small scaling factor, which is adsorbed into the rescaled
temperature $\beta'$. In the next section, we will discuss the steps
leading to Eq. (\ref{alpha}) in more detail.

The model defined in Eqs. (\ref{fluorescence}) and (\ref{alpha}) contains
the four fitting parameters $A$, $\beta$, $\beta'$ and $\tilde{c}$ which
were fitted against the spike-in data of the Affymetrix array HGU95a in
Ref. \cite{carl06}.  The parameters $\beta'$, $\tilde{c}$ and $A$ will
be discussed in Sec. \ref{sec:hyb_sol} and Sec. \ref{sec:saturation}.
The parameter $\beta$ is the inverse temperature. Instead of fixing
it to the experimental value we have kept it as a fitting parameter
as explained in Ref. \cite{carl06}. The hybridization free energies
$\Delta G$ and $\Delta G_R$ are calculated from tabulated experimental
data for DNA/RNA \cite{sugi95_sh,sugi00} and RNA/RNA \cite{xia98_sh}
duplex formation in solution.

We note that we fit mismatches and perfect matches with the same model.
The difference between the two is that there is a different hybridization
free energy $\Delta G$: one expects a lower signal for mismatches compared
to perfect matches, due to weaker binding. This is not always the case;
as remarked in several studies for a substantial fraction of probes (30\%,
as reported in Ref. \cite{naef03}) one observes ``bright mismatches''
for which the mismatch intensity $I_{\rm MM}$ exceeds the intensity
$I_{\rm PM}$ of the perfect match. However, it has been observed
\cite{bind05} that bright MM come predominantly from probes with low
intensity, which suggests that bright mismatches are associated with
weak specific hybridization when the signal $I$ is dominated by $I_0$
in Eq. (\ref{fluorescence}).

In recent work \cite{heim06} we also compared the current model
with the approach based on position-dependent effective affinities as
for instance described in Refs. \cite{naef03,bind05}. The conclusion is
that the two approaches are fully consistent with each other, provided
that various effects are incorporated such as partial unzipping of
the probe-target complex, less than 100\% efficiency in the probe
growth during lithography, and entropic repulsion between the target
and the substrate.  These additional effects are the main factors
causing position-dependence (and thus allowing for a comparison with
position-dependent effective affinities); for a quantitative prediction
of the intensities, their combined effect can be well approximated by
a slight decrease of $\beta$ in Eq. (\ref{fluorescence}) and they are
therefore not included in the current study.

\section{On the hybridization in solution}
\label{sec:hyb_sol}

We now discuss the approximations leading to the form of $\alpha$.
We denote the concentration of free 25-mer targets in solution as $[t]$,
the concentration of free target strands that are complementary from
nucleotide $i$ up and including nucleotide $j$ as $[\hat{t}_{i,j}]$, and
the concentration of duplexes between these two as $[t\,\hat{t}_{i,j}]$.
Chemical equilibrium (see Fig. \ref{FIG00}(b)) yields for the equilibrium
constant:
\be
K_{i,j} = \frac{[t][\hat{t}_{i,j}]} {[t\hat{t}_{i,j}]} = e^{-\beta \Delta G_R (i,j)},
\label{equilib}
\ee
where $\Delta G_R (i,j)$ is the RNA/RNA hybridization free energy for
target molecules in solution, which are complementary from nucleotide $i$ up
and including $j$, and $\beta=1.59$ mol/kcal (corresponding to the experimental
temperature of 45 degrees).
For a given gene, the measure of the
gene expression level which one wants to determine is the total target
concentration $c$ given by
\be
c=[t] + \sum_{i,j} [t \hat{t}_{i,j}].
\label{conserv}
\ee
Solving Eqs.(\ref{conserv}) and (\ref{equilib}) we find
for the fraction of single stranded target in solution:
\be
\alpha_f = \frac{[t]}{c} = \frac{1}{1+\sum_{i,j} [\hat{t}_{i,j}] 
\exp (\beta \Delta G_R (i,j)) }.
\label{alpha_full}
\ee
Note that the summation in the denominator of Eq. (\ref{alpha_full})
was replaced in the approximate expression Eq. (\ref{alpha}) by
the single term $\tilde{c}\exp(\beta' \Delta G_R^{(37)})$, with
fitting parameters $\tilde{c}$ and $\beta'$.

\begin{figure}
\includegraphics[width=8.0cm]{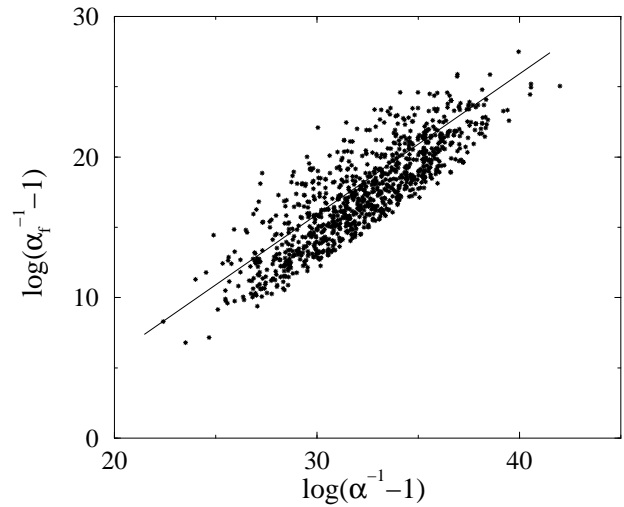}
\caption{Comparison of the summation in Eq.~(\ref{alpha_full}), equal
to $\alpha_f^{-1}-1$, and its approximation in Eq.~\ref{alpha}),
equal to $\alpha^{-1}-1$, for the first 1,000 spike-in sequences of
HGU133.  Note that a change in $c_0$ corresponds to a vertical shift
over $\log(c_0)$; in this figure, we used $c_0=1$.  The straight line
is a fit, given by $y=x+b$ with $b=-14.1$.
\label{alphacompare}
}
\end{figure}

Eq.~(\ref{alpha_full}) requires as input estimates of the
concentration $[\hat{t}_{i,j}]$ of complementary sequences with length
$l=j-i+1$, present in solution.  Assuming that all four nucleotides
are roughly equally abundant, and that there are no correlations along
the sequence, the abundance of short sequences with length $l$ will
decrease as $[\hat{t}_{i,j}] \sim 4^{-l}$.  This scaling breaks down
beyond some length $L$; assuming for the human transscriptome a total
length of $10^7$ nucleotides, a random sequence longer than 12 is more
likely not present at all, since $4^{12} \gtrsim 10^7$. We therefore
take as our approximation
\be
[\hat{t}_{i,j}] = 
\left\{ 
\begin{array}{cc}
c_0\cdot 4^{-(j-i)}	& {\rm for \ j-i< 12,}\\
0		        & {\rm otherwise.}
\end{array} 
\right.
\label{concdrop}
\ee
Here, $c_0$ is a measure of the RNA concentration.  Using this
approximation for the concentration of complementary strands, we can now
compare Eqs.~(\ref{alpha}) and (\ref{alpha_full}).  Fig. \ref{alphacompare}
shows the more elaborate model Eq.(\ref{alpha_full}) as a function of
the approximate form Eq.~(\ref{alpha}), with the values for the fitting
parameters $\beta'$ and $\tilde{c}$ taken from Ref.~\cite{carl06}.
There is a reasonable agreement between the two.

%%%%%%%%%%%%%%%%%%%%%%%%%%%%%%%%% FIG_01 %%%%%%%%%%%%%%%%%%%%%%%%%%%%%%%%%%%
%%%%%%%%%%%%%%%%%%%%%%%%%%%%%%%%% FIG_01 %%%%%%%%%%%%%%%%%%%%%%%%%%%%%%%%%%%

Since Eq.~(\ref{alpha_full}) has a better microscopic foundation than
Eq.~(\ref{alpha}), it should in principle allow for a better estimate
of the hybridization in solution.  There are however severe limitations
to the use of Eq.~(\ref{alpha_full}).
In the hybridization in solution, there is a competition between the
contributions of short sequences, which are abundant but have a low
affinity, versus long sequences, for which the concentration is low but
the affinity high. The concentration drops on average approximately by a
factor of 4 per added length (see Eq.~(\ref{concdrop})), but the affinity
grows by approximately $\langle \Delta G\rangle \approx$ 2 or 3 kcal/mol,
the average value of RNA/RNA interaction parameters~\cite{bloo00}. Since
$\exp(\beta \langle \Delta G\rangle) > 4$, the longer sequences dominate
the hybridization in solution. However, as discussed above, beyond length
$L\approx 12$, there simply are no complementary strands. The accuracy
of the more elaborate model Eq.~(\ref{alpha_full}) thus hinges crucially
on knowing the longest complementary strand which is transcribed, as
well as its affinity and its concentration.  Since the approximate model
Eq.~(ref{alpha}) is not expected to perform worse than the more elaborate
model Eq.~(\ref{alpha_full}), we keep using the former.

The data points in Fig.~\ref{alphacompare} can be fitted by a
straight line with slope 1: the value of $\beta'=0.67$ mol/kcal
in Ref.\cite{carl06}, corresponding to 725 K, apparently is the
appropriate value to describe the experiments at a temperature
of 45 degrees. The offset in the straight-line fit is equal to
$\log(\tilde{c})-\log(c_0)$. Since the straight-line fit has an offset of
-14.1, and since we used the fitted value of $\tilde{c}=2\cdot 10^{-2}
pM$ in Ref.~\cite{carl06}, an estimate of the RNA concentration is
$c_0=\exp(14.1)\cdot \tilde{c}=30$ nM.  Even if we do not use the more
elaborate model Eq.~(\ref{alpha_full}), it provides us with a microscopic
basis for the values of the parameters $\beta'$ and $\tilde{c}$ in the
approximate model Eq.~(\ref{alpha}).

%%%%%%%%%%%%%%%%%%%%%%%%%%%%%%%%% FIG_01 %%%%%%%%%%%%%%%%%%%%%%%%%%%%%%%%%%%
\begin{figure}[t]
\includegraphics[width=7.0cm]{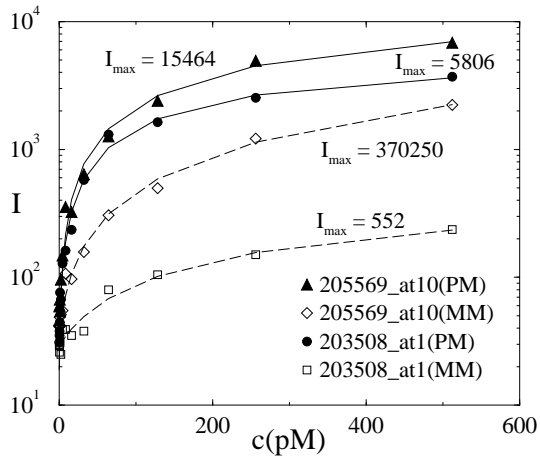}
\caption{Plot on intensity vs. concentration for three spike-in genes
of the HGU133 chipset. $I_{\rm max}$ indicates the saturation value 
obtained from a three parameters ($I_0$, $A$ and $K$) non-linear fit 
based on Eq. (\ref{fit_c}).}
\label{Ivsc}
\end{figure}
%%%%%%%%%%%%%%%%%%%%%%%%%%%%%%%%% FIG_01 %%%%%%%%%%%%%%%%%%%%%%%%%%%%%%%%%%%

\section{On the signal saturation level}
\label{sec:saturation}

If the target concentration $c$ and the binding energy $\Delta G$
are sufficiently high, the Langmuir isotherm saturates to a maximal
value. From Eq. (\ref{fluorescence}) we find for
$c \exp(\beta \Delta G) \gg 1$
\be
I_{\rm max} = I_0 + A \approx A,
\ee
where we have used the fact that typically the background level, $I_0$,
is much lower than the value of $A$. The saturation intensity arises if
targets are bound to almost all probes. Since the number of probes does
not vary between the sequences being measured, this saturation intensity
is also expected to be sequence-independent, and more specifically,
should not distinguish between perfect matches and mismatches.  A recent
analysis of the Latin square set \cite{held03,burd04} reported widely
different values for the saturation intensity. It is worth clarifying
further this issue here.

The obvious procedure to determine the saturation intensity, is to look at
the intensity of a probe as a function of concentration. Assuming an
effective affinity $K_s$ for probe sequence $s$, the intensity $I_s(c)$ as a
function of concentration $c$ is given by
\be
I_s(c) = I_{0,s} + \frac{A_s c K_s}{1+c K_s},
\label{fit_c}
\ee
in which $I_{0,s}$ is the (sequence-dependent) background intensity
due to non-specific binding.  A plot of $I_s$ vs. $c$ for two probes of
the HGU133 spike-in set is shown in Fig. \ref{Ivsc}. Taking $I_0$, $A$
and $K$ in eq.~(\ref{fit_c}) as fitting parameters, and extrapolating
to high concentration then yields the saturation intensity.

%%%%%%%%%%%%%%%%%%%%%%%%%%%%%%%%% FIG_01 %%%%%%%%%%%%%%%%%%%%%%%%%%%%%%%%%%%
\begin{figure}[t]
\includegraphics[width=8.0cm]{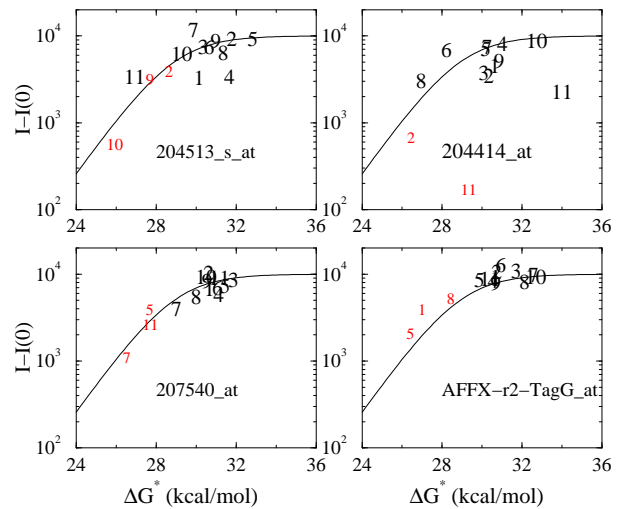}
\caption{Plot of $I-I_0$ as a function of $\Delta G - R T\log \alpha$
for 4 sequences spiked-in at a concentration of $c=512$ pM.  The numbers
indicate the probe set numbers. Smaller characters are used for the
MM signals. Solid lines represent the Langmuir model as given by
Eq. (\ref{alpha}). The data are consistent, except few outliers, with 
the Langmuir model with roughly constant saturation level $A \approx 10^4$.}
\label{IvsDG_star}
\end{figure}
%%%%%%%%%%%%%%%%%%%%%%%%%%%%%%%%% FIG_01 %%%%%%%%%%%%%%%%%%%%%%%%%%%%%%%%%%%

%%%%%%%%%%%%%%%%%%%%%%%%%%%%%%%%% FIG_01 %%%%%%%%%%%%%%%%%%%%%%%%%%%%%%%%%%%
\begin{figure*}[t]
\includegraphics[width=7.5cm]{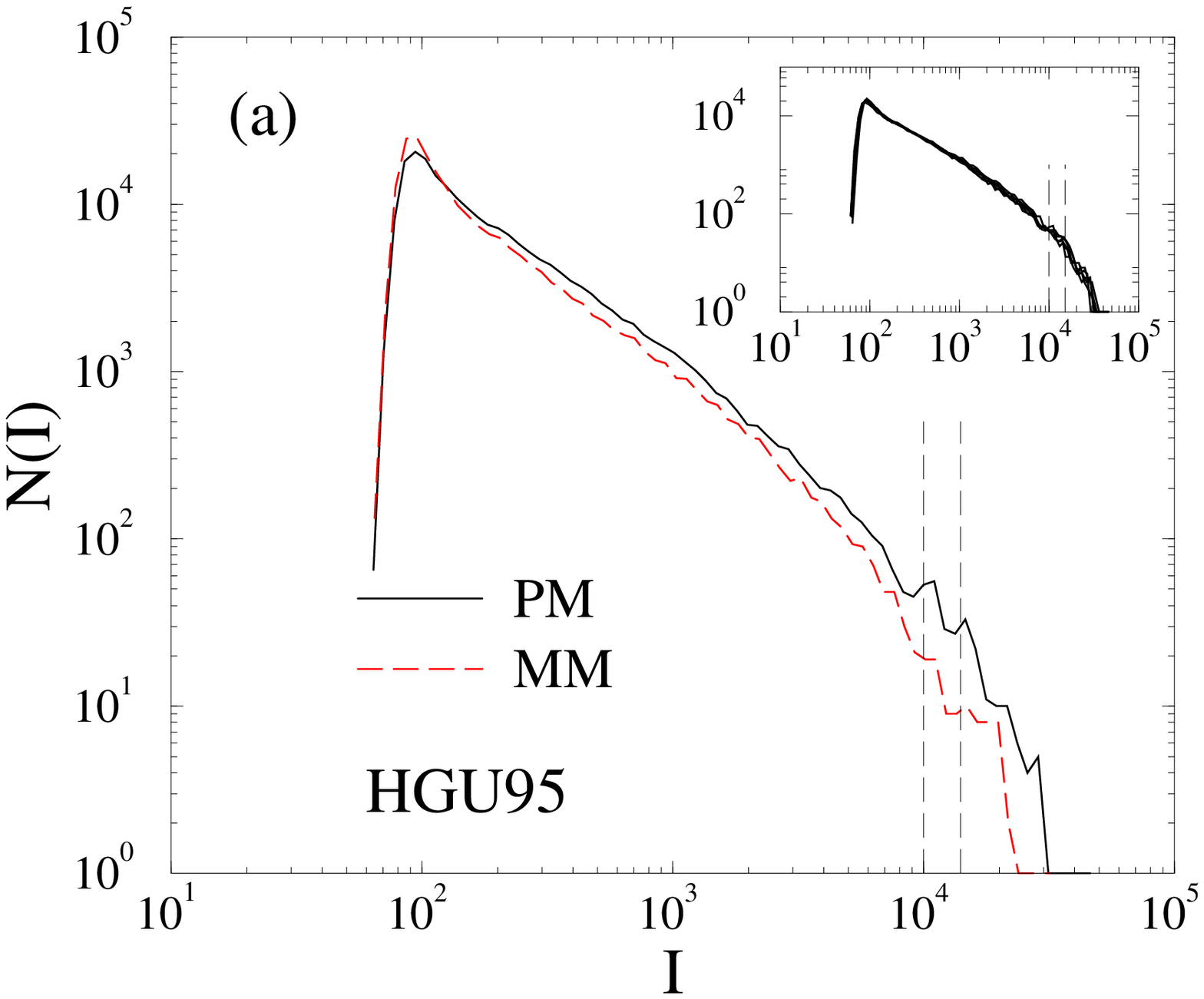}
\includegraphics[width=7.5cm]{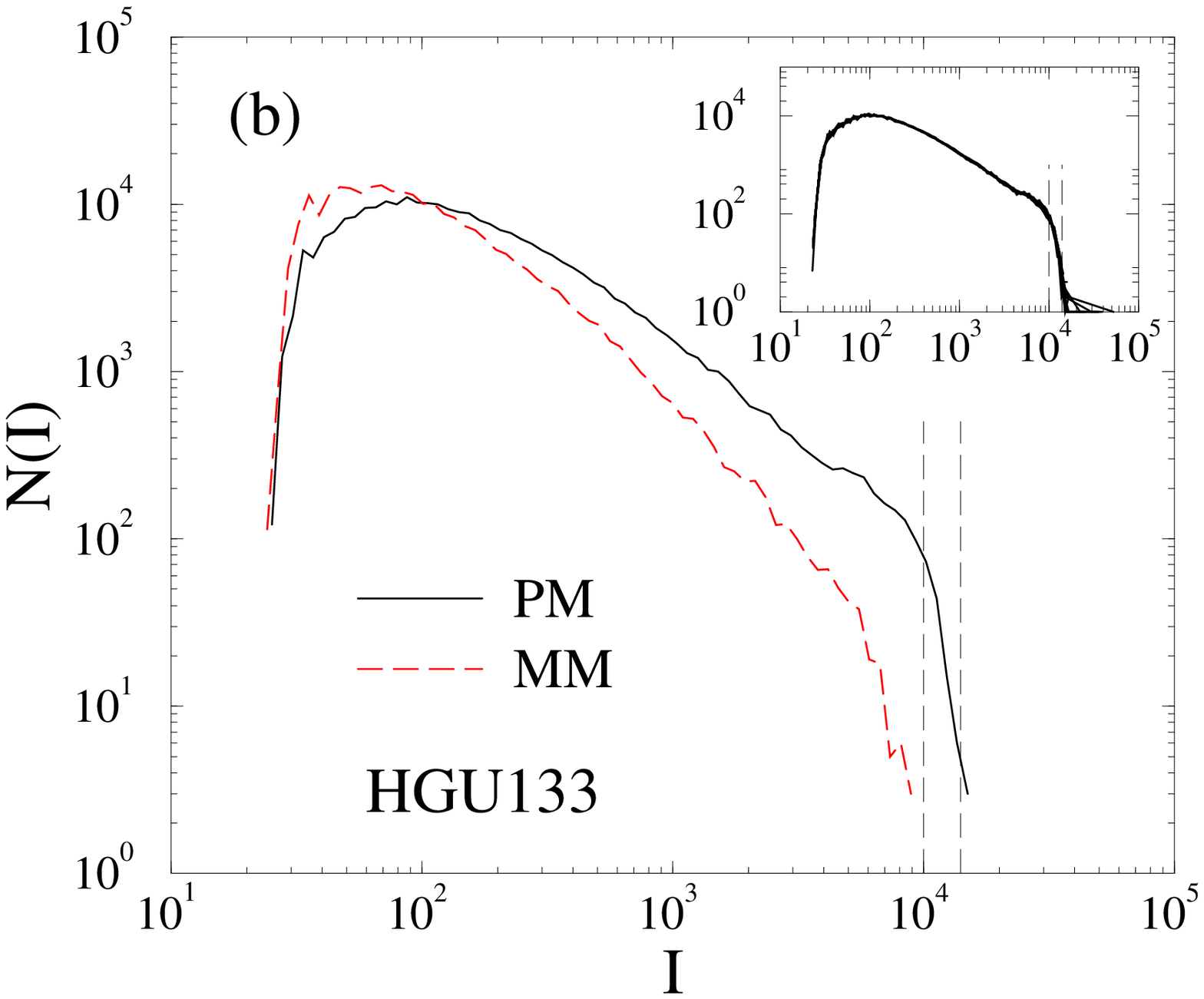}
\caption{Histograms of the PM and MM intensities for the Latin square
experiments in log-log scale for the chips HGU95a (a) and HGU133 (b). The
plots contain 19 histograms referring to different experiments (a) and
12 experiments (b).  The dashed lines are positioned at $I=10000$ and
$I=15000$ (intensities are given in Affymetrix scale). Insets: histograms
of the total intensity of PM and MM together.}
\label{FIG0h}
\end{figure*}
%%%%%%%%%%%%%%%%%%%%%%%%%%%%%%%%% FIG_01 %%%%%%%%%%%%%%%%%%%%%%%%%%%%%%%%%%%

Two research groups \cite{held03,burd04} followed this procedure, and both
found saturation intensities that vary wildly between different sequences.
A first effect that can cause deviations from the Langmuir fit
Eq.~(\ref{fit_c}) is that the lithographic process, through which
the probes are synthesized in situ in Affymetrix chips, is not 100\%
efficient. As estimated by Burden~\cite{burd04}, only about 10\% of the
probes reach the full length of 25 nucleotides. At low intensities far
from saturation, the incomplete probes can be safely ignored since their
affinity is much lower than that of the fully grown probes. However, under
conditions where the fully grown probes are saturated, clearly there will
be contributions to the fluorescent intensity from the almost complete
probes, and an even further increase in concentration will bring into
play shorter and shorter incomplete probes.  Consequently, the Langmuir
fit Eq.~(\ref{fit_c}) breaks down near saturation; extrapolation to high
concentration is an unreliable procedure.

A second cause of worry is that comparing fluorescent intensities from
different chips is also potentially unreliable, since the microarrays
might have undergone slightly different processing during the washing
and staining. Since Affymetrix microarrays cannot be reused, the
spike-in measurements used in Refs.~\cite{held03,burd04} required a new
chip for each concentration.

To avoid these two potential sources of error, we therefore consider
the intensities for a given probe set at a specific concentration,
i.e. $c$ constant and $\Delta G$ and $\alpha$ variables in
Eq. (\ref{fluorescence}).  The data belong to the same array.
An example of this type of analysis is shown in Fig. \ref{IvsDG_star}
for a concentration of $c=512$ pM.  On the horizontal axis we plot
$\Delta G^* = \Delta G - RT \log \alpha$.  The solid lines are
given by the Langmuir curve Eq. (\ref{fluorescence}).  Note that
the large majority of the probes align along the expected curve,
with few exceptions as for instance probe 11 (both PM and MM) for
the probe set 204414\_at.  Therefore, the data are consistent with
a value of $A$ roughly constant in Eq. (\ref{fluorescence}), which
suggests indeed that the large variations in $I_{\rm max}$ obtained
from the extrapolations of the data in the earlier analysis are more
likely to be an artifact of the extrapolations. Note however that
some variability of the saturation level can be seen in the data of
Fig. \ref{IvsDG_star}. Typically this variability is of about $20\%$. In
order to keep the model simple we will keep $A$ constant in the rest of
the paper. An interesting possible explanation of the variability of $A$
has been given in Ref. \cite{burd04}, i.e. that this variation is due
to the post-hybridization washing of the array.

Yet another different way of addressing the issue of the saturation
intensities is to analyze the histogram of the intensities on the whole
chip, as in Fig. \ref{FIG0h}, which shows both the intensities for the
HGU95 and HGU133 spike-in data. To reveal the data at high intensities,
they are plotted in a log-log scale. In the figure we note a drop in the
histogram around $I \approx 10\ 000$, sharper in the HGU133 chipset,
which is consistent with the estimate of the saturation intensity
obtained from the fits of intensities vs $\Delta G - R T \log \alpha$,
as given in Fig. \ref{IvsDG_star}. Note that in Fig. \ref{FIG0h}(b)
the drop is 100-fold in the range $10\ 000 < I < 15\ 000$, which
suggests that the data are consistent with a roughly constant value
of the saturation. However a more close inspection of the histogram
of the HGU133 for PM and MM intensities separately, reveals that the
estimated saturation value for the two may be different. In the case of
PM intensities alone the drop is rather sharp at around $I \approx 10\
000$, however the MM intensities seem to saturate at lower intensities,
which is not seen in the HGU95 data (Fig. \ref{FIG0h}(a)).  The number
of MM probes reaching an intensity close to the saturation level in the
histogram of Fig. \ref{FIG0h}(b) is quite small so the fact that the the
MM and PM reach a different saturation level cannot be concluded for sure.

Also the low-intensity side of the histograms in Fig.~\ref{FIG0h} contain
interesting information. Both for the HGU95 and HGU133, the intensity
drops steeply below a minimal intensity. For HGU95, this drop occurs
around $I_{\rm min}\approx 70$, while for HGU133 the drop occurs around
$I_{\rm min}\approx 30$. This increase of the dynamical intensity range
by more than a factor of two is a clear demonstration of the fast rate
of improvement in microarray technology.

%%%%%%%%%%%%%%%%%%%%%%%%%%%%%%%%% FIG_01 %%%%%%%%%%%%%%%%%%%%%%%%%%%%%%%%%%%
\begin{figure*}[t]
\includegraphics[width=4.2cm]{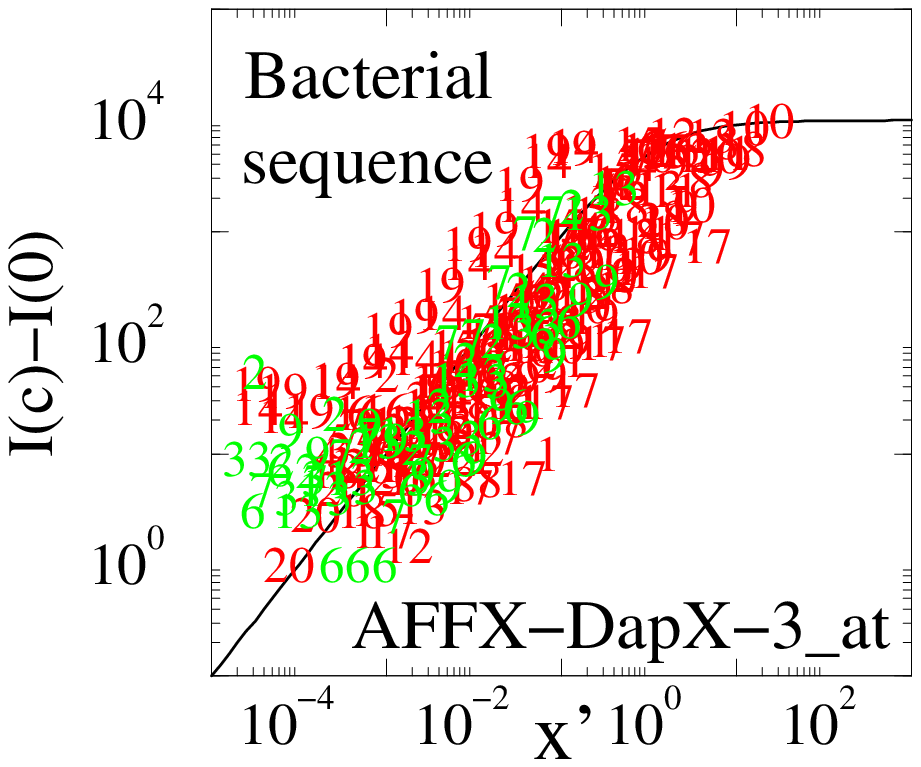}
\includegraphics[width=4.2cm]{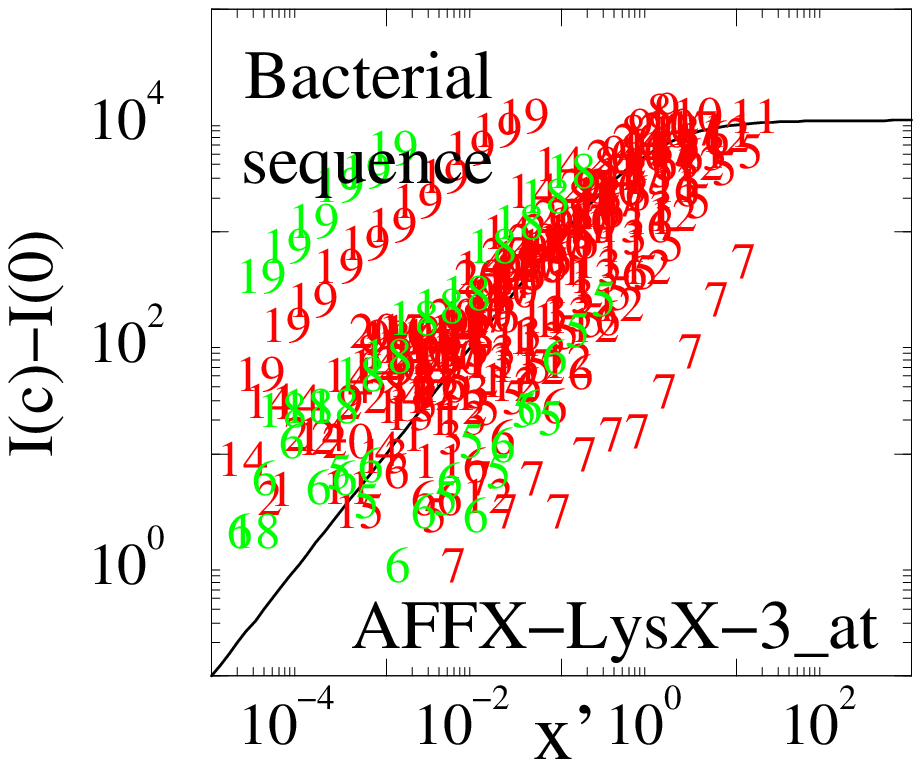}
\includegraphics[width=4.2cm]{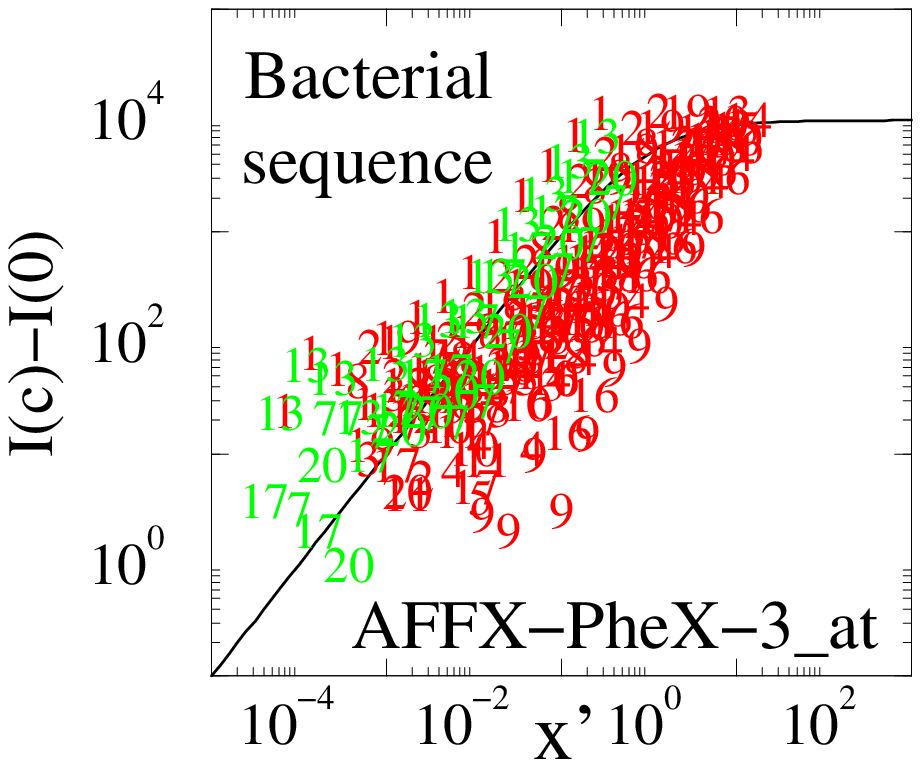}
\includegraphics[width=4.2cm]{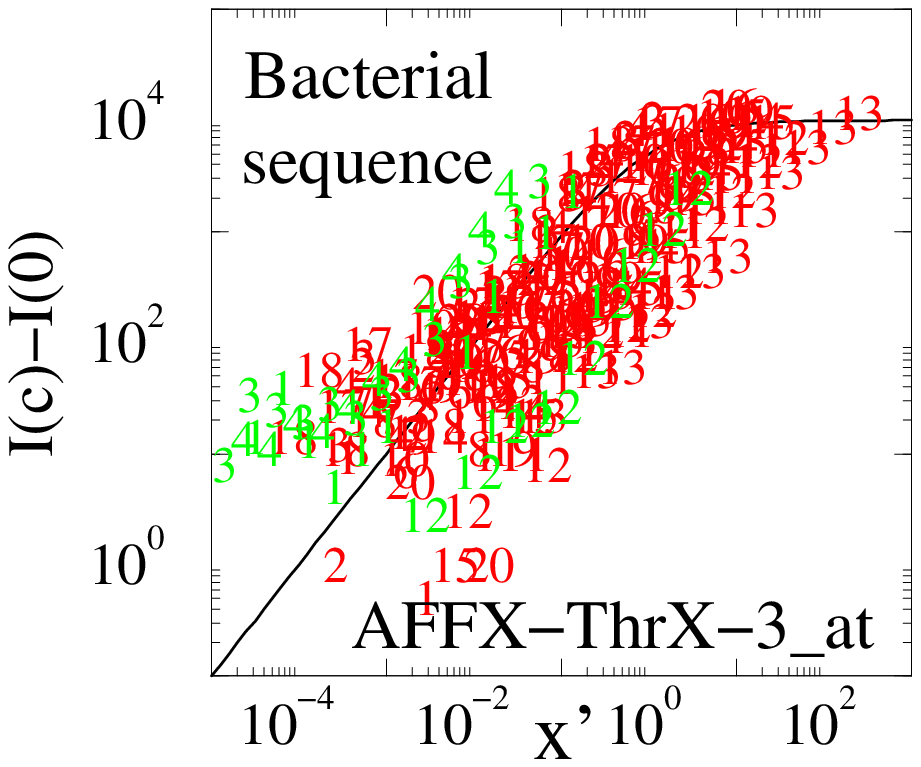}

\includegraphics[width=4.2cm]{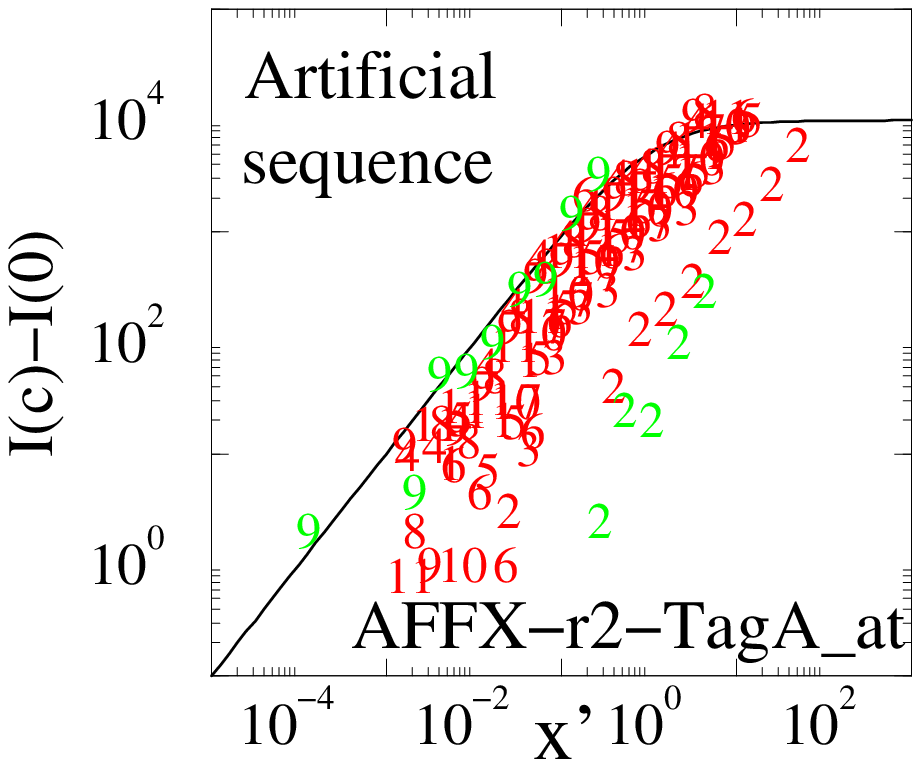}
\includegraphics[width=4.2cm]{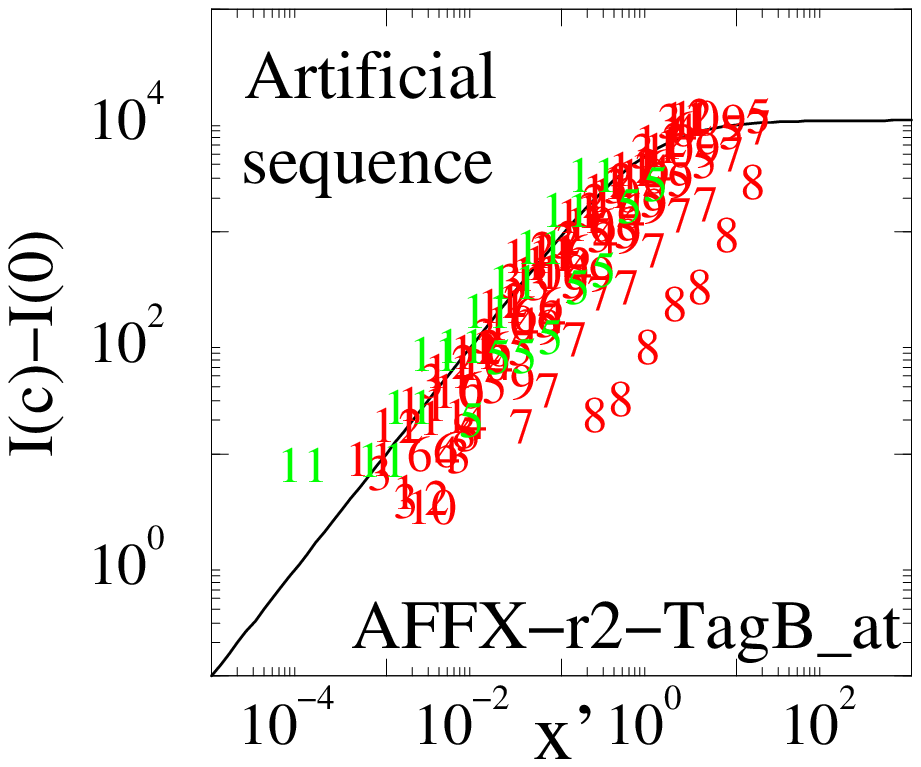}
\includegraphics[width=4.2cm]{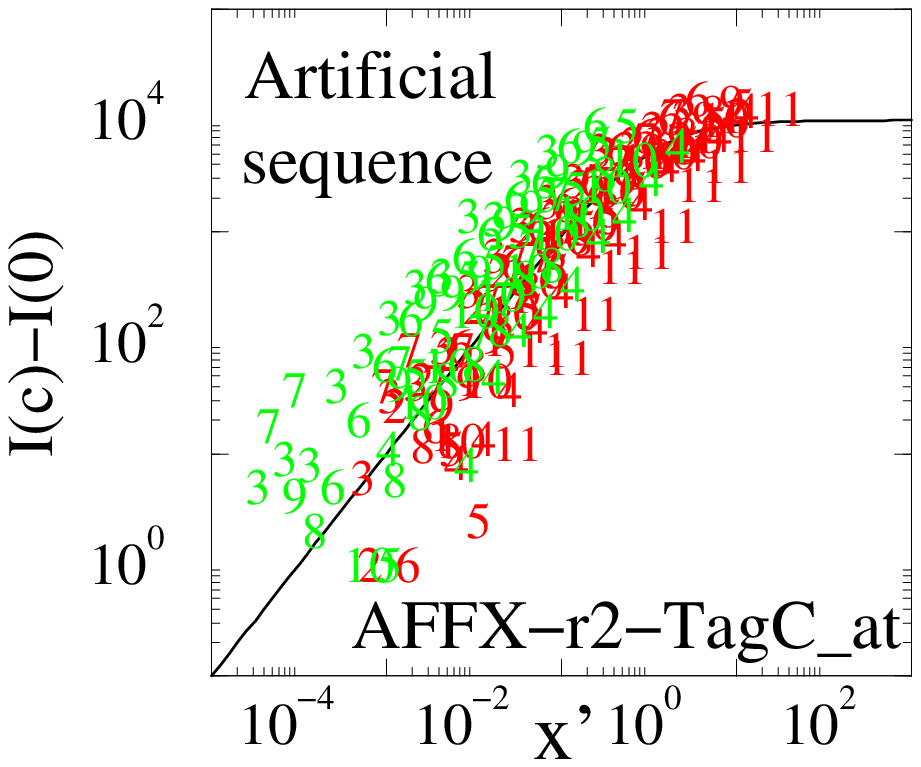}
\includegraphics[width=4.2cm]{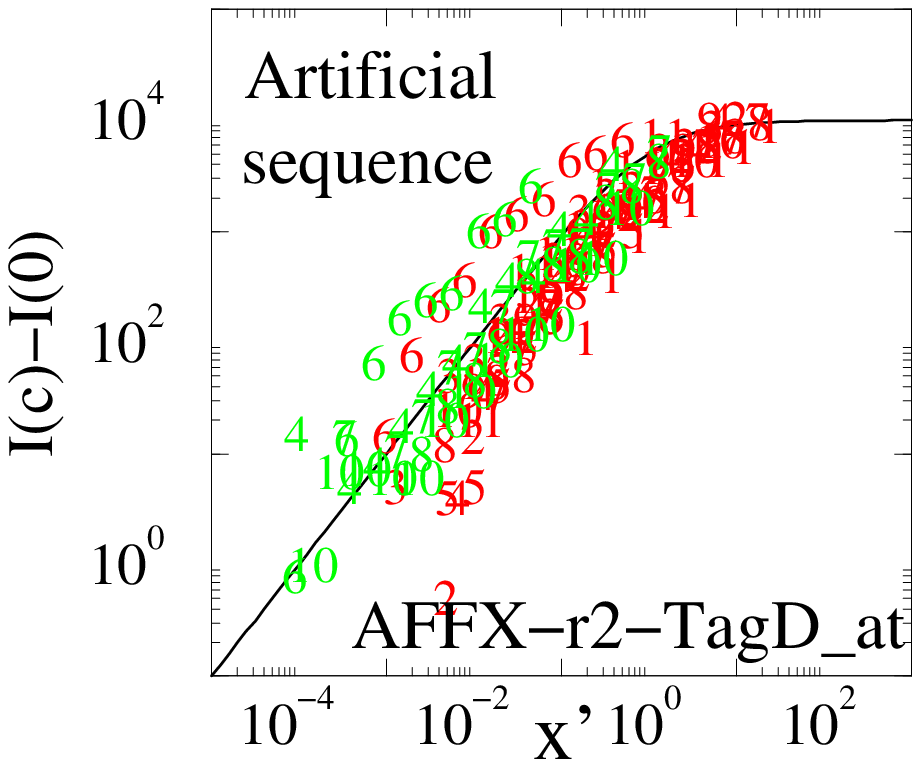}

\includegraphics[width=4.2cm]{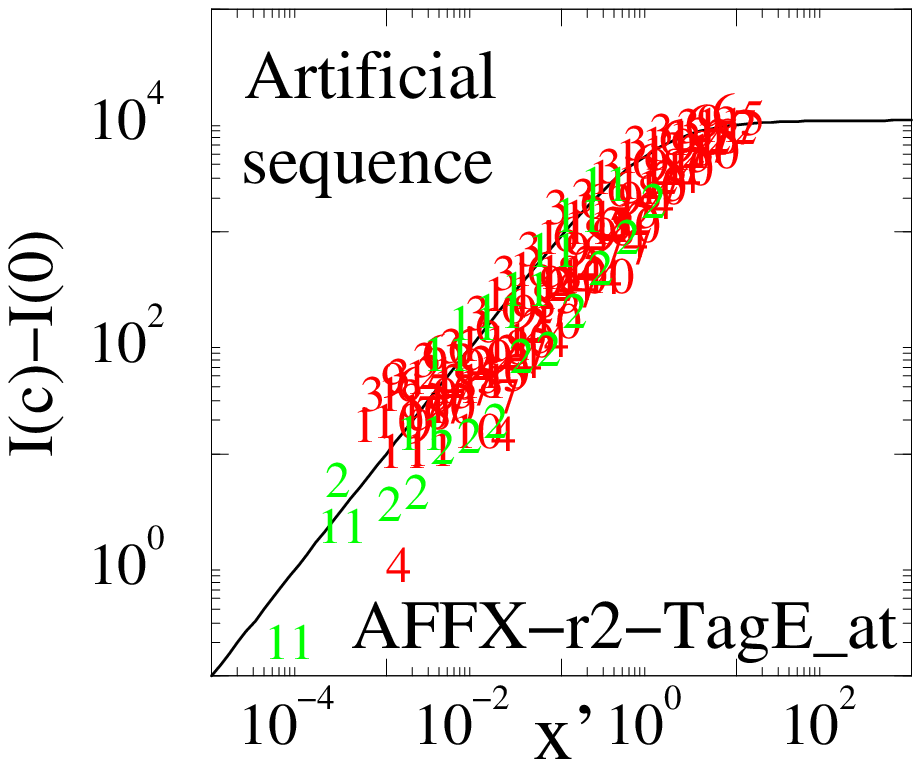}
\includegraphics[width=4.2cm]{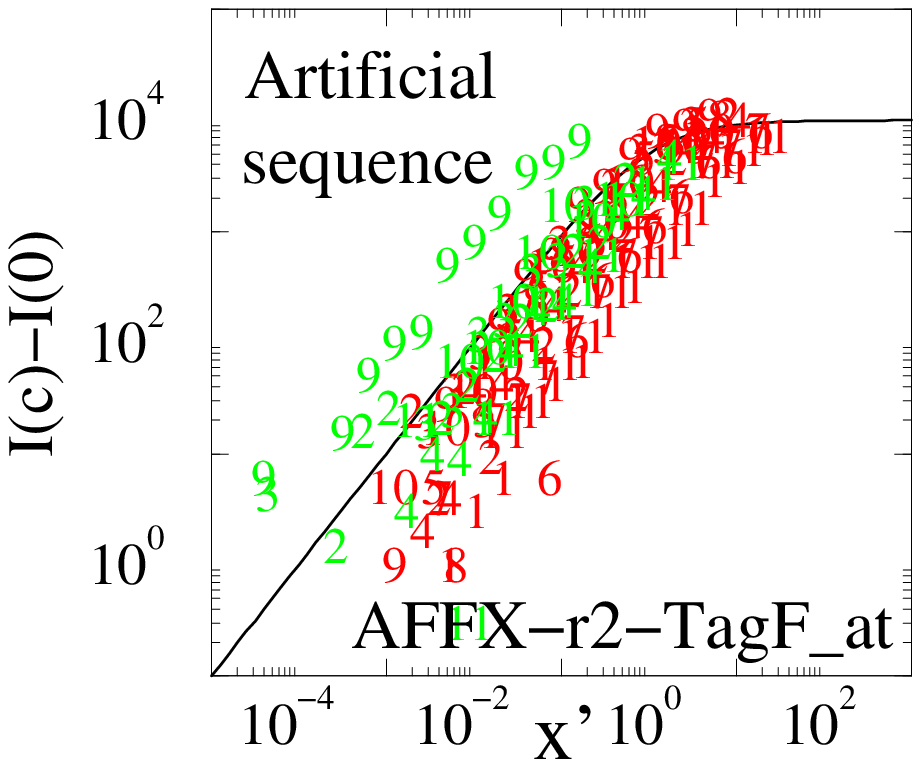}
\includegraphics[width=4.2cm]{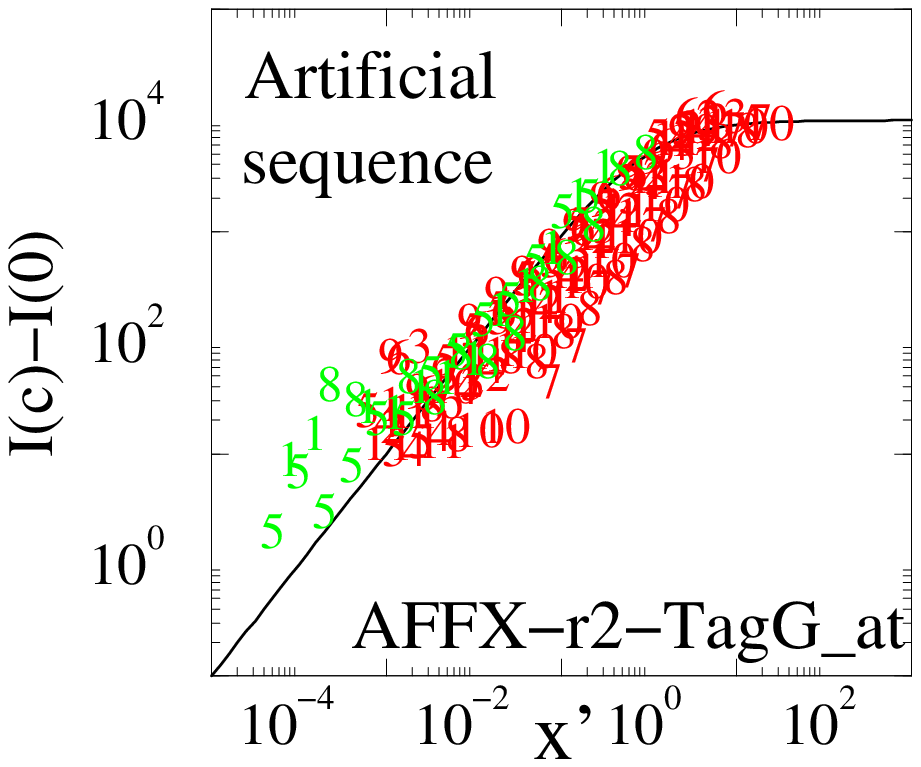}
\includegraphics[width=4.2cm]{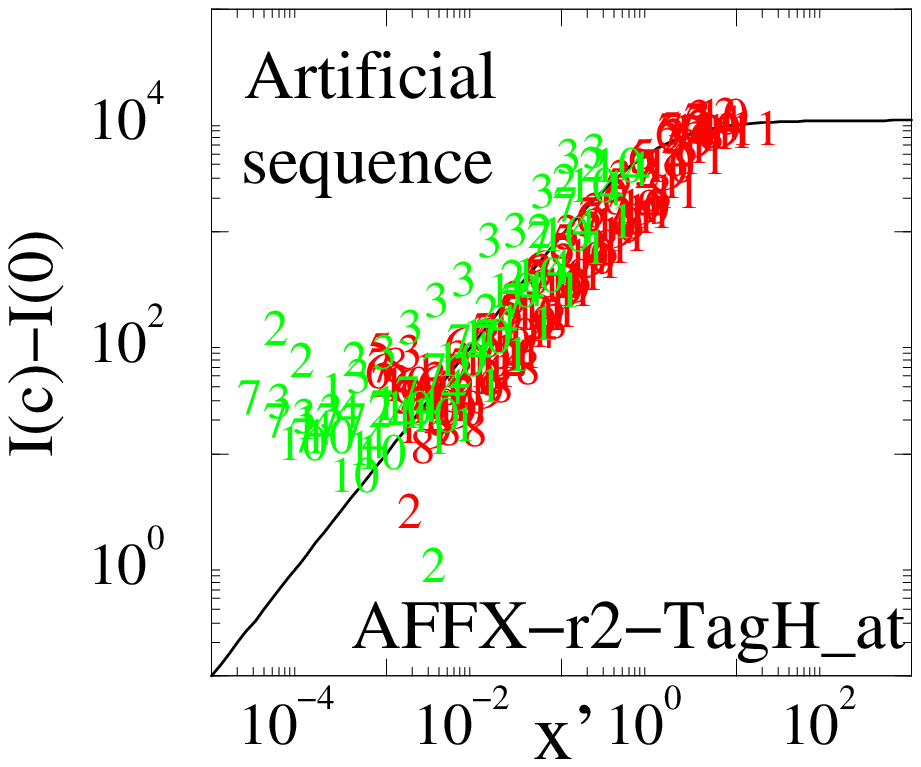}
\caption{Collapse plots for the 4 bacterial and the 8 artificial 
sequences of the HGU133 spike-in set.
In these plots the background subtracted intensities for a given probe
set are plotted as functions of the rescaled variable $x'$ given in 
Eq.~(\ref{xprime}). The data corresponds to all spike-in concentrations
for a given probe sets. Solid lines correspond to the Langmuir isotherm.
Compared with the human and bacterial sequences the 
artificial sequences are characterized by the best collapses.}
% \caption{Collapse plots for the Artificial sequences in the HGU133 
% spike-in set. Compared with the human and bacterial sequences the 
% artificial sequences are characterized by the best collapses.}
\label{collapse_A}
\end{figure*}
%%%%%%%%%%%%%%%%%%%%%%%%%%%%%%%%% FIG_01 %%%%%%%%%%%%%%%%%%%%%%%%%%%%%%%%%%%

\section{Analysis of data collapses}
\label{sec:analysis}

As a test of the validity of the model we plotted \cite{carl06} the
data as a function of the rescaled variable:
\be
x' = \alpha c e^{\beta \Delta G}.
\label{xprime}
\ee
If the model is to be trusted the data for different values of $c$ and
different probe sequences (i.e. different $\Delta G$ and $\alpha$)
ought to ``collapse" onto a single master curve
\be
I - I_0 = \frac{A x'}{1+x'}.
\label{rescaled}
\ee
This collapse has indeed been observed in the large majority of the
spike-in genes of the HGU95a chipset \cite{carl06}. Interestingly, the
very few outliers observed in that case could be explained as annotation
errors or unbalance of free energies used for specific nucleotides,
as discussed in Ref. \cite{carl06}.

We choose here the same fitting parameters used in Ref. \cite{carl06}
for the HGU95 chipset, that is: $A= 10\ 000$, $\beta = 0.74$ mol/kcal,
$\beta'= 0.67$ mol/kcal and $\tilde{c}= 10^{-2}$ pM. These parameters
fit equally well the HGU133 spike-in data.

%%%%%%%%%%%%%%%%%%%%%%%%%%%%%%%%% FIG_01 %%%%%%%%%%%%%%%%%%%%%%%%%%%%%%%%%%%
\begin{figure*}[t]
\includegraphics[width=4.2cm]{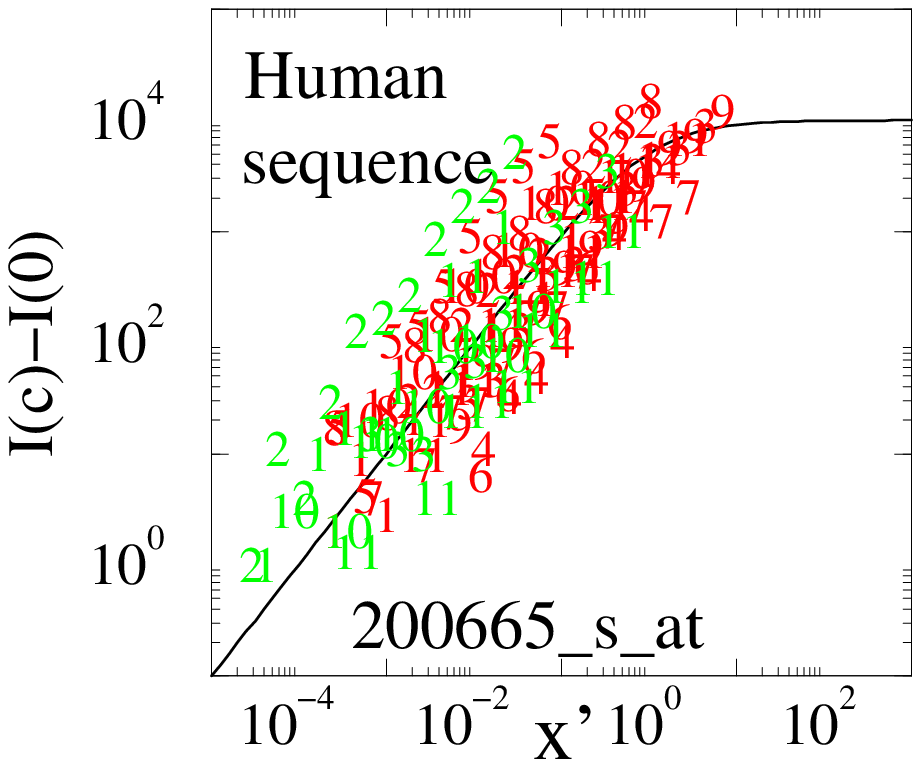}
\includegraphics[width=4.2cm]{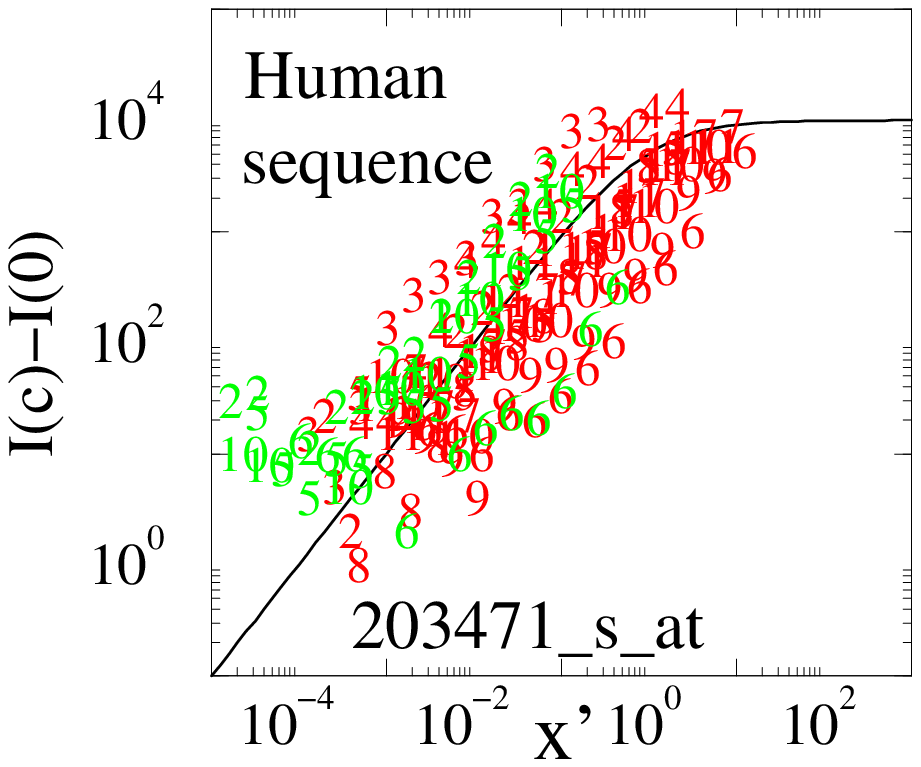}
\includegraphics[width=4.2cm]{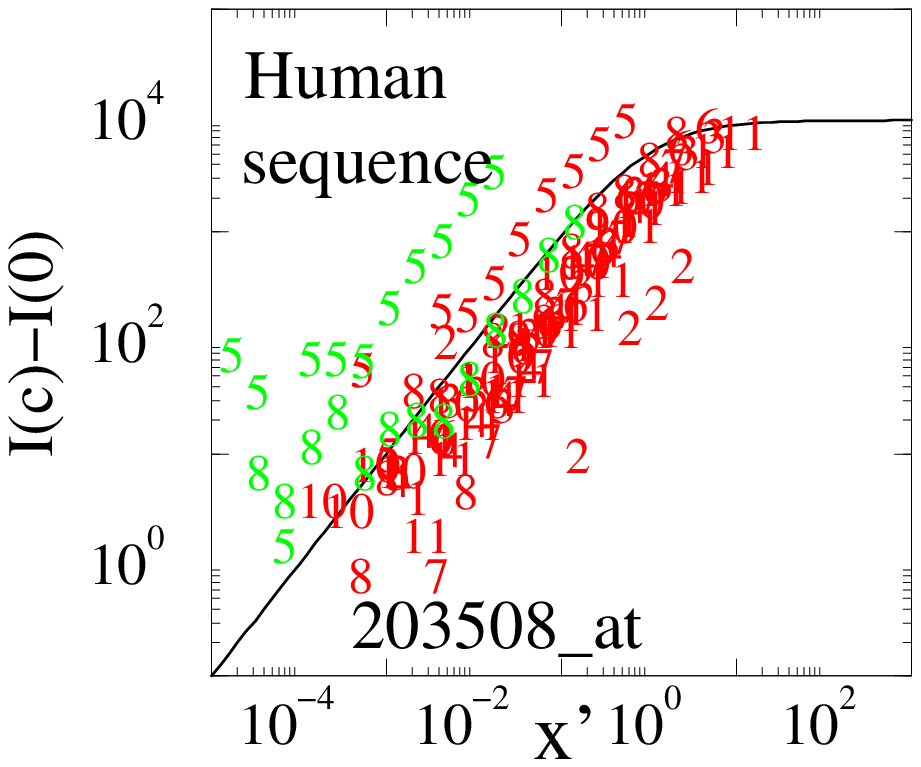}
\includegraphics[width=4.2cm]{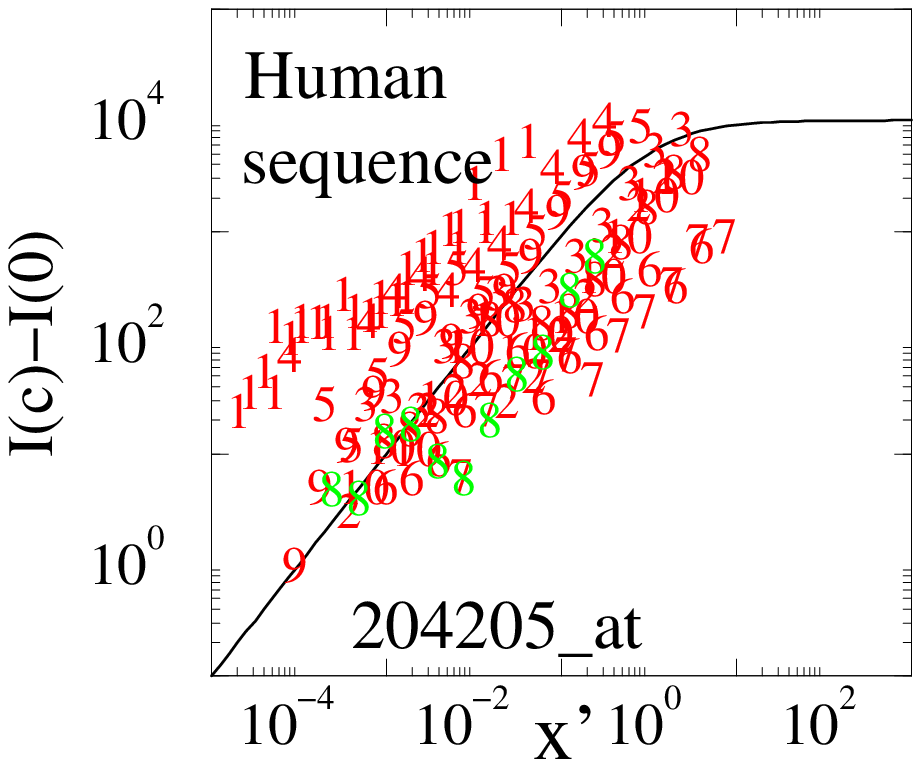}

\includegraphics[width=4.2cm]{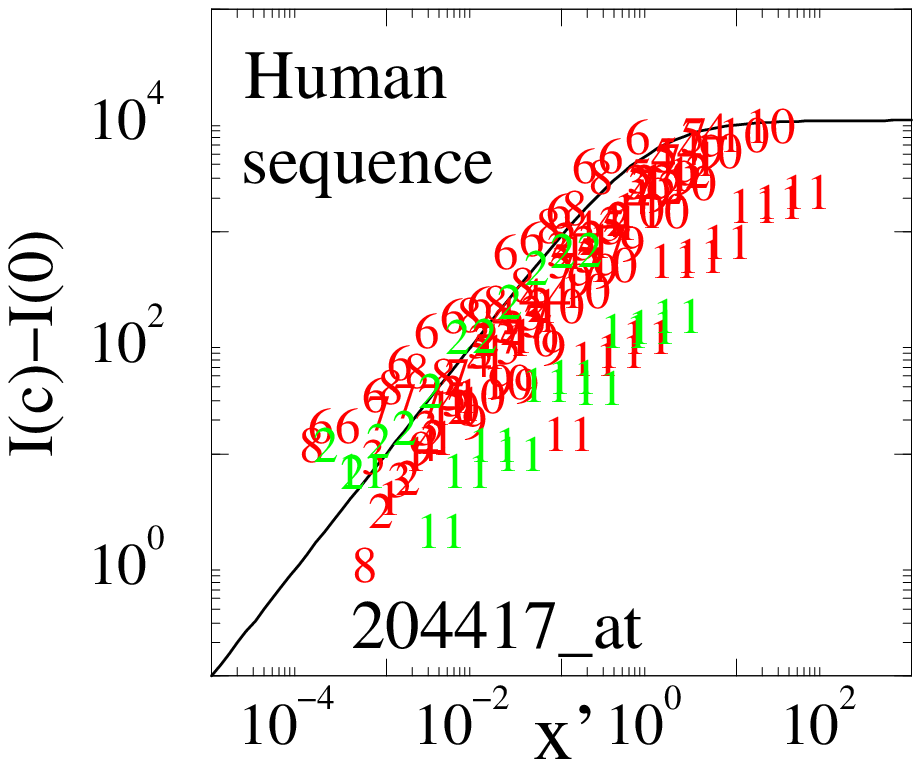}
\includegraphics[width=4.2cm]{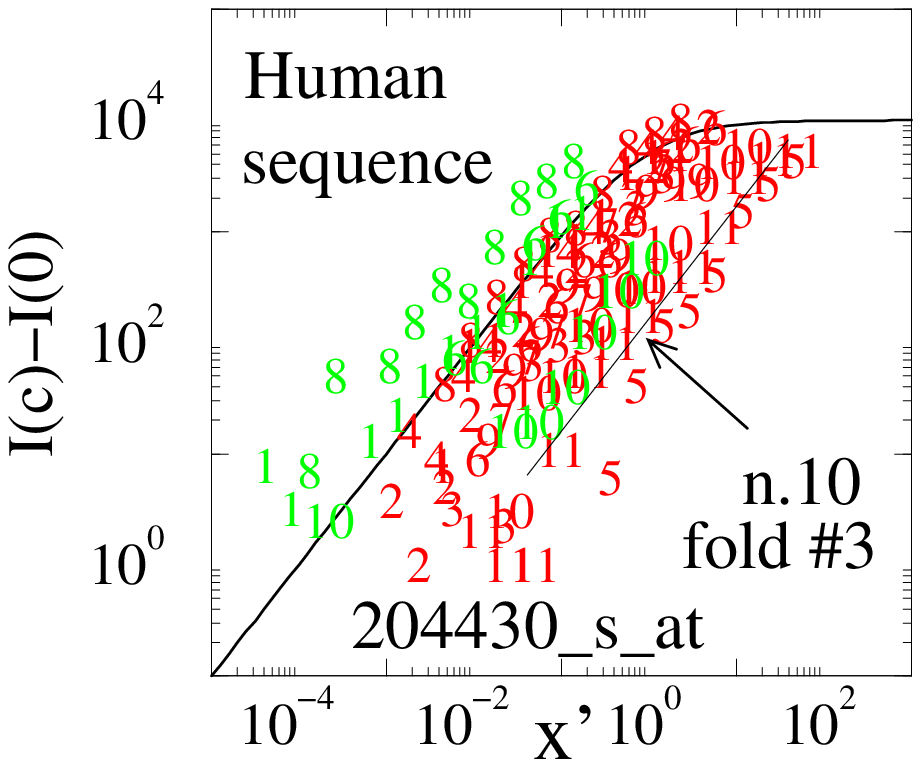}
\includegraphics[width=4.2cm]{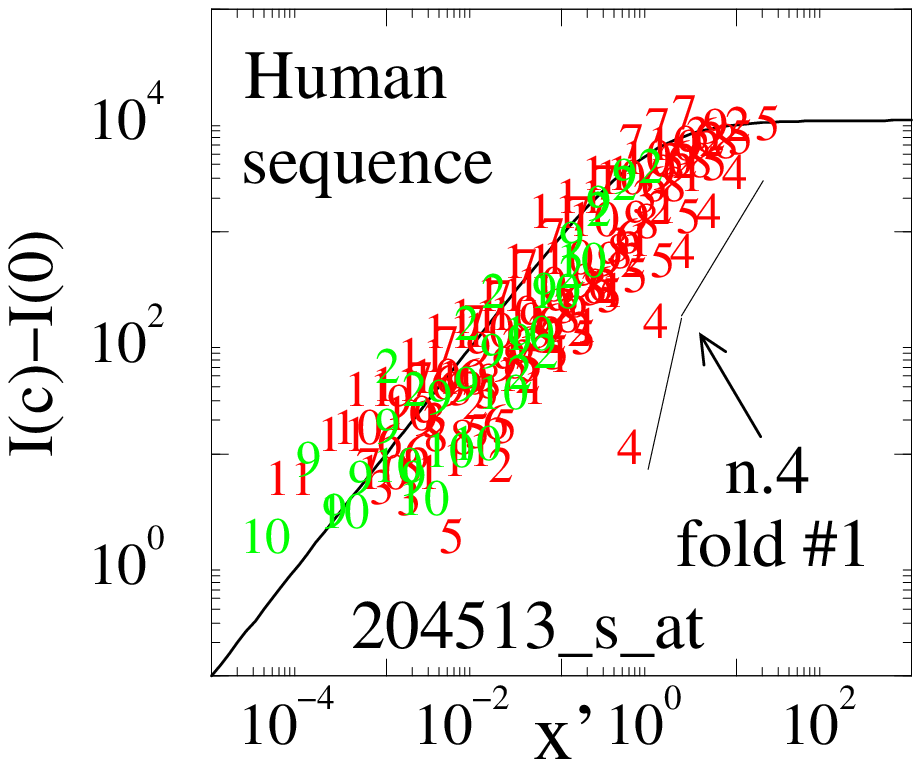}
\includegraphics[width=4.2cm]{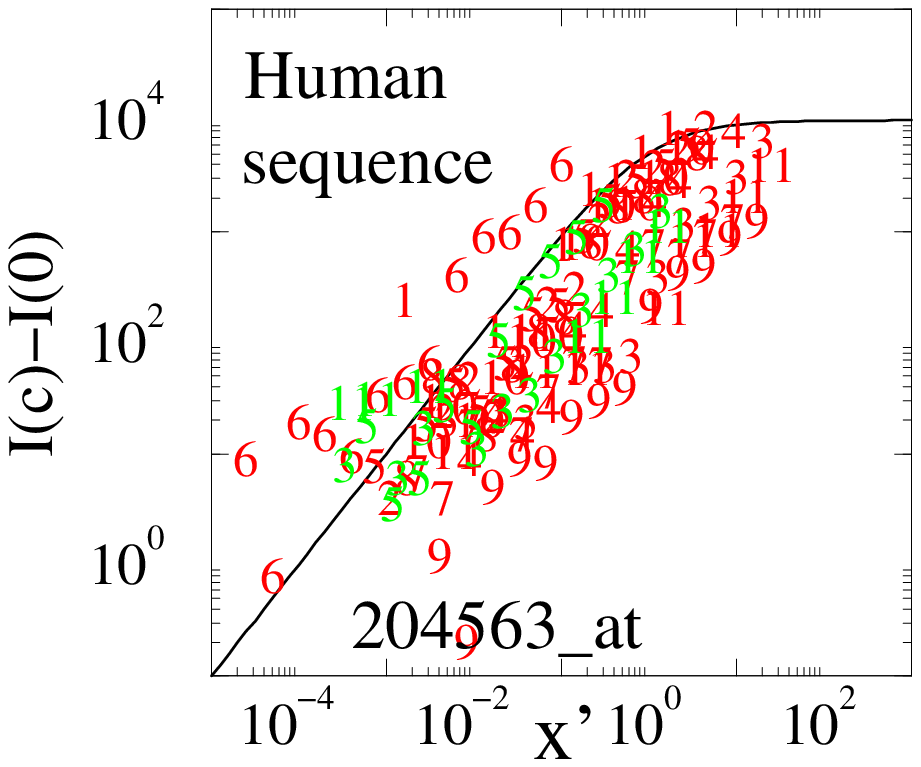}

\includegraphics[width=4.2cm]{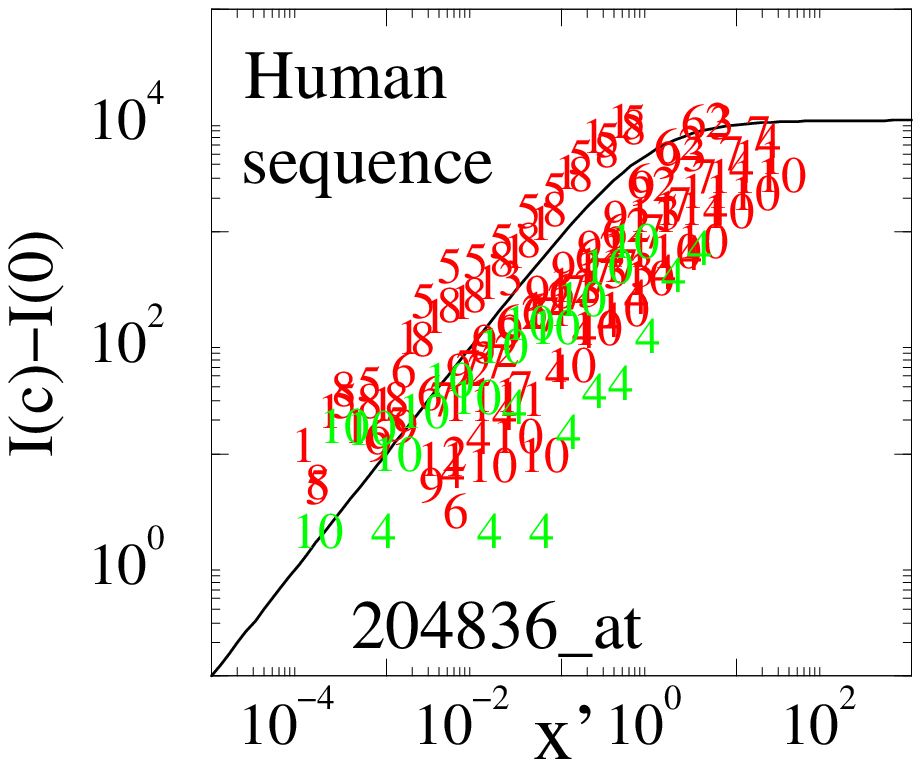}
\includegraphics[width=4.2cm]{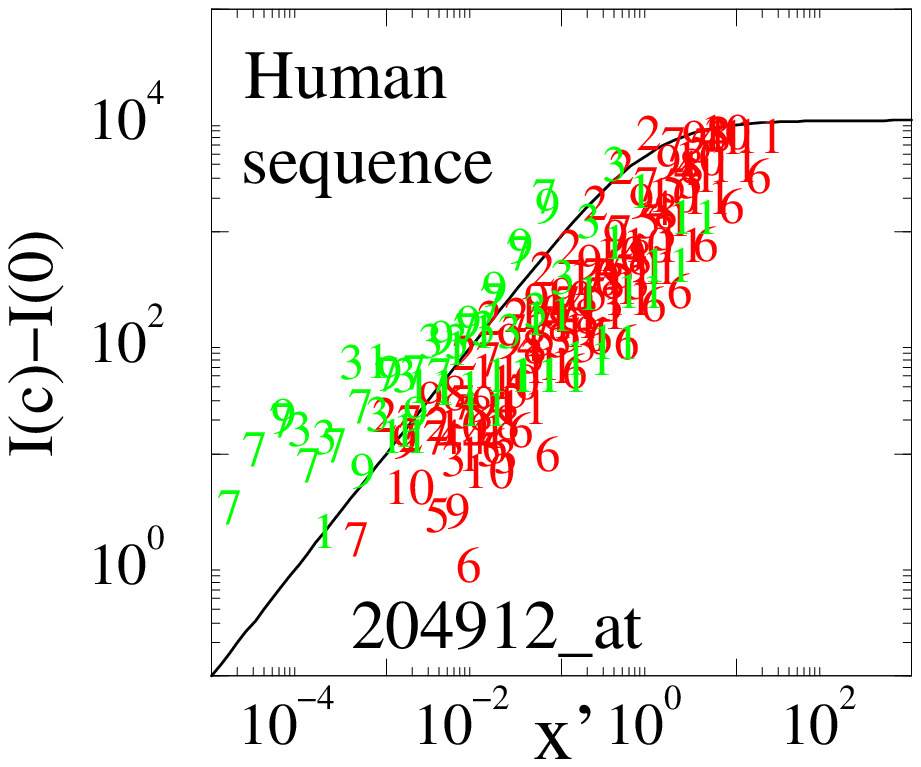}
\includegraphics[width=4.2cm]{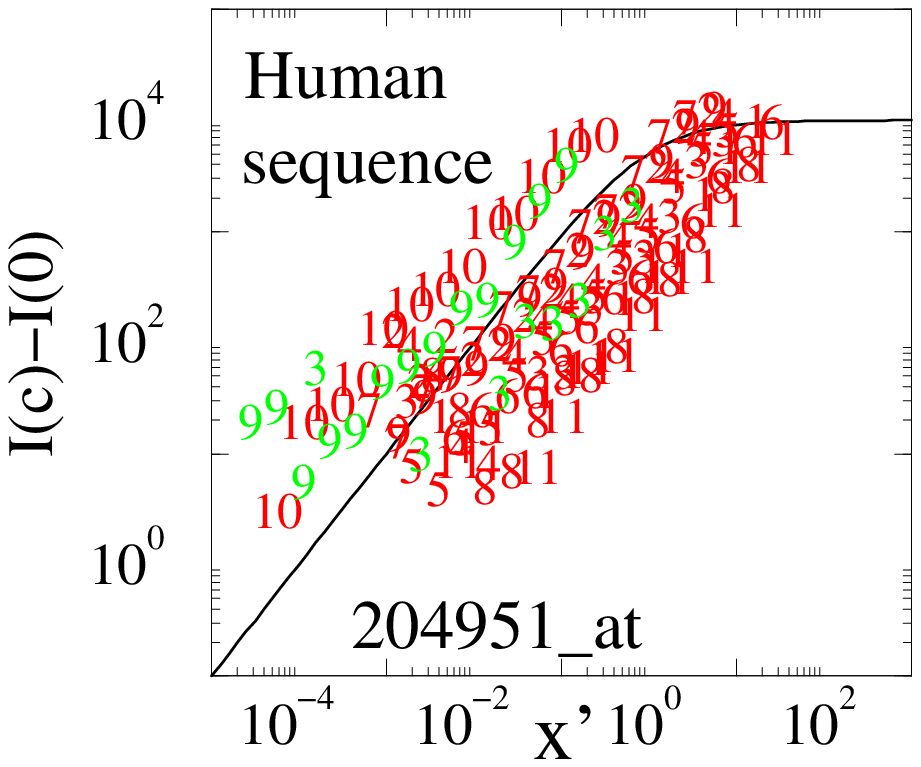}
\includegraphics[width=4.2cm]{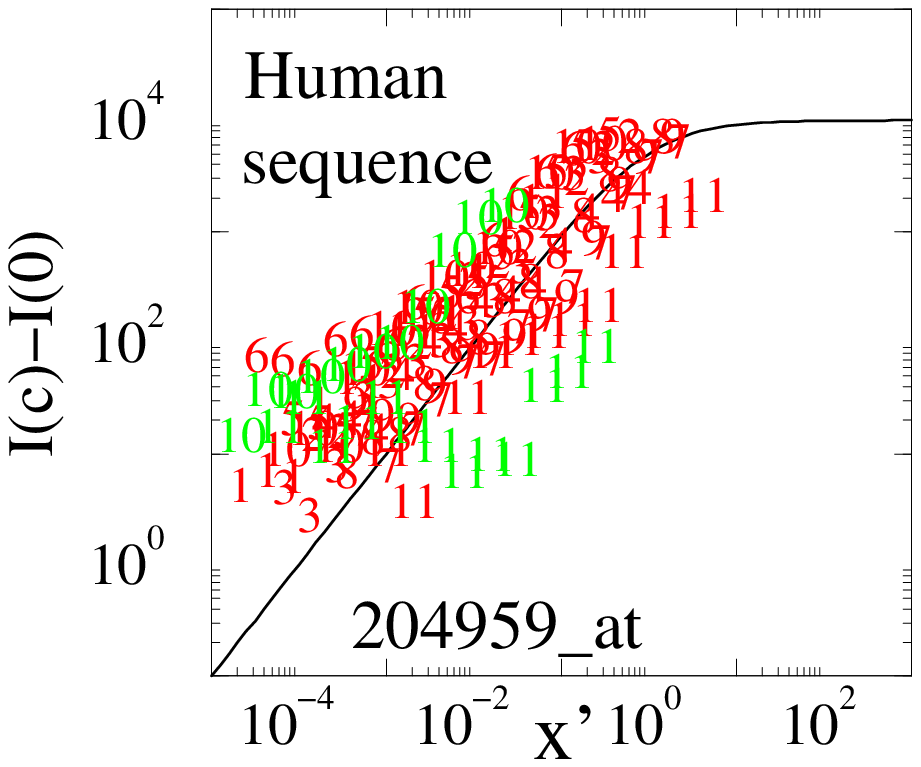}

\includegraphics[width=4.2cm]{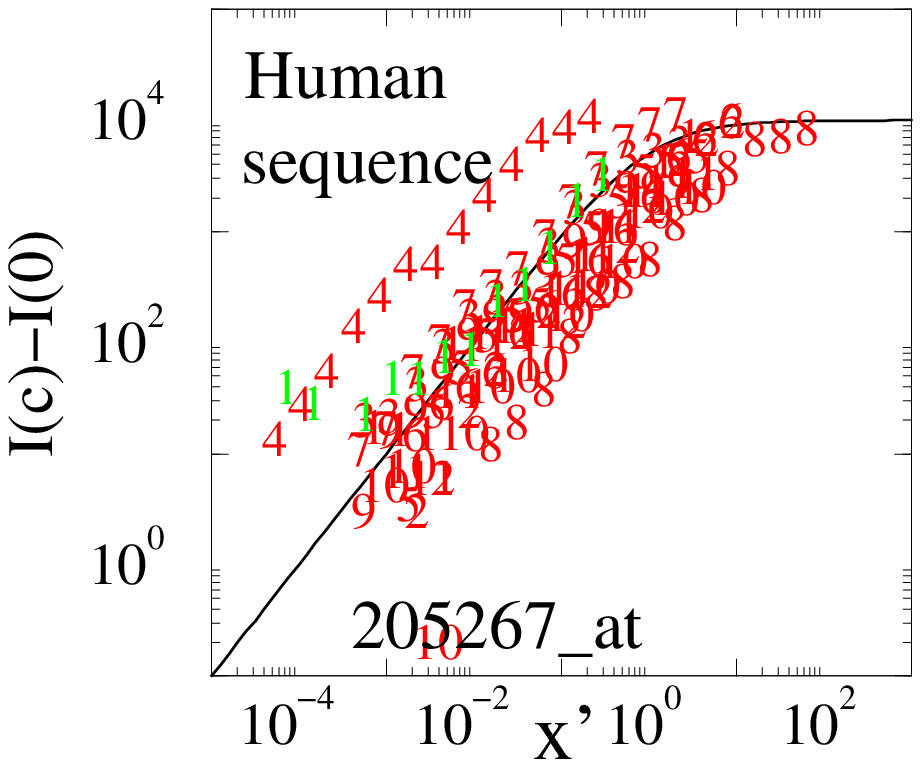}
\includegraphics[width=4.2cm]{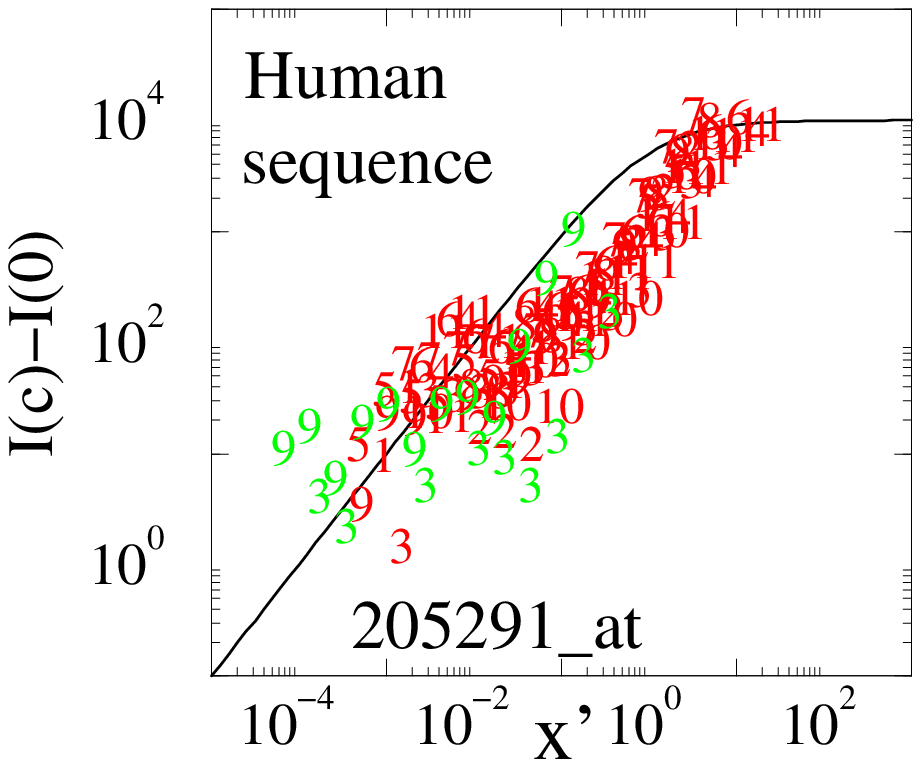}
\caption{Collapse plots for Human sequences of the HGU133 spike-in set
(part 1). The probes which are complementary to targets which the largest
folding free energies are emphasized (see Table \ref{table_fold}). They
correspond to probes 204912\_at10 and 204513\_s\_at4.
}
\label{collapse_H1}
\end{figure*}
%%%%%%%%%%%%%%%%%%%%%%%%%%%%%%%%% FIG_01 %%%%%%%%%%%%%%%%%%%%%%%%%%%%%%%%%%%

%%%%%%%%%%%%%%%%%%%%%%%%%%%%%%%%%%%%%%%%%%%%%%%%%%%%%%%%%%%%%%%%%%%%%%%%%%%%%%%%
\begin{table}[b]
\caption{List of values of $\langle w \rangle$ and $\sigma_w$ for the bacterial 
and the artificial sequences in the spike-in set HGU133.}
\begin{ruledtabular}
\begin{tabular}{lll|lll}
Probe set & $\langle w \rangle$ & $\sigma_w$ & Probe set & $\langle w \rangle$ & $\sigma_w$\\
\hline
AFFX-DapX-3\_at & 0.08 & 1.49 & AFFX-PheX-3\_at     & 0.16  & 1.55\\
AFFX-LysX-3\_at & 0.89 & 2.46 & AFFX-ThrX-3\_at     & 0.22  & 1.59\\
AFFX-r2-TagA\_at  & -1.05 & 0.97 & AFFX-r2-TagE\_at & -0.32 & 0.82\\
AFFX-r2-TagB\_at  & -0.51 & 0.83 & AFFX-r2-TagF\_at & -0.46 & 1.09\\
AFFX-r2-TagC\_at  &  0.43 & 1.08 & AFFX-r2-TagG\_at & -0.11 & 0.90\\
AFFX-r2-TagD\_at  & -0.03 & 0.90 & AFFX-r2-TagH\_at &  0.11 & 1.22
\end{tabular}
\end{ruledtabular}
\label{table_bacterial}
\end{table}

%%%%%%%%%%%%%%%%%%%%%%%%%%%%%%%%%%%%%%%%%%%%%%%%%%%%%%%%%%%%%%%%%%%%%%%%%%%%%%%%
\begin{table}[t]
\caption{List of values of $\langle w \rangle$ and $\sigma_w$ for 
the human sequences in the spike-in set HGU133.}
\begin{ruledtabular}
\begin{tabular}{lcc|lcc}
Probe set & $\langle w \rangle$ & $\sigma_w$ & Probe set & $\langle w \rangle$ & $\sigma_w$\\
\hline
200665\_s\_at   &  0.54 & 1.26   & 205569\_at         &  -0.28 & 1.12\\
203471\_s\_at   &  0.39 & 1.43   & 205692\_s\_at      &   0.24 & 1.27\\
203508\_at      &  0.45 & 1.83   & 205790\_at         &  -0.78 & 0.76\\
204205\_at      &  0.86 & 2.11   & 206060\_s\_at      &   0.52 & 1.66\\
204417\_at      & -0.24 & 1.18   & 207160\_at         &  -0.32 & 1.06\\
204430\_s\_at   & -0.48 & 1.13   & 207540\_s\_at      &  -0.29 & 0.62\\
204513\_s\_at   & -0.68 & 1.16   & 207641\_at         &   0.24 & 2.72\\
204563\_at      & -0.57 & 1.44   & 207655\_s\_at      &   0.76 & 1.06\\
204836\_at      & -0.04 & 1.41   & 207777\_s\_at      &  -0.14 & 1.11\\
204912\_at      & -0.31 & 1.35   & 207968\_s\_at      &  -0.85 & 1.66\\
204951\_at      & -0.15 & 1.48   & 209354\_at         &   0.04 & 1.41\\
204959\_at      &  1.33 & 1.62   & 209606\_at         &   0.77 & 1.44\\
205267\_at      &  0.36 & 1.23   & 209734\_at         &  -0.20 & 1.51\\
205291\_at      & -0.44 & 1.24   & 209795\_at         &   0.63 & 1.71\\
205398\_s\_at   & -0.15 & 1.37   & 212827\_at         &   0.61 & 2.53
\end{tabular}
\end{ruledtabular}
\label{table_human}
\end{table}
%%%%%%%%%%%%%%%%%%%%%%%%%%%%%%%%%%%%%%%%%%%%%%%%%%%%%%%%%%%%%%%%%%%%%%%%%%%%%%%%

\begin{figure*}[t]
\includegraphics[width=4.2cm]{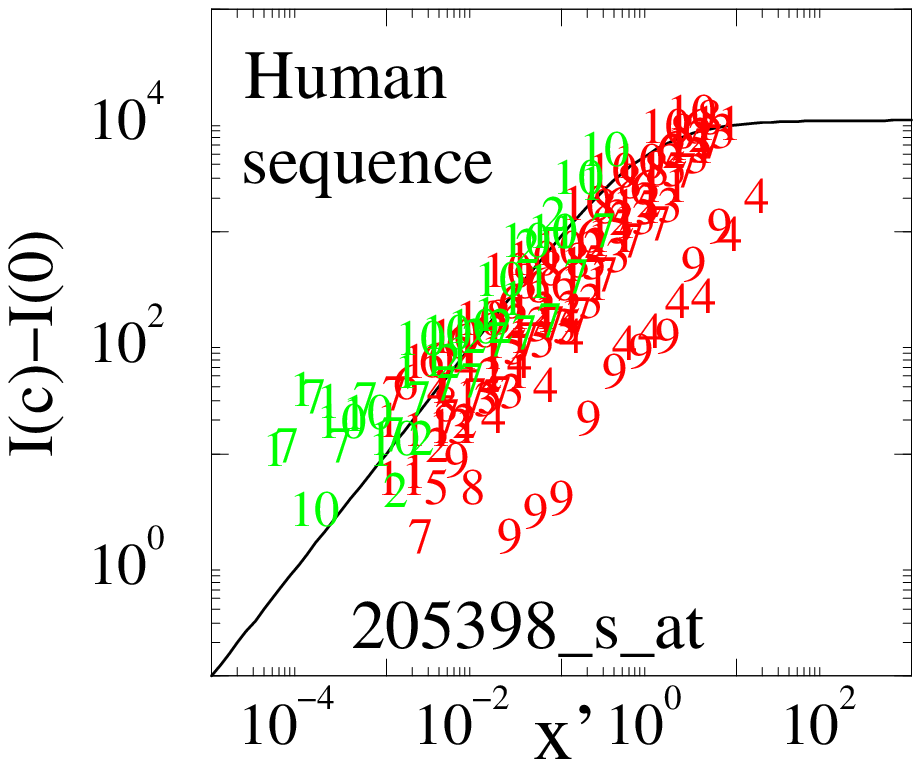}
\includegraphics[width=4.2cm]{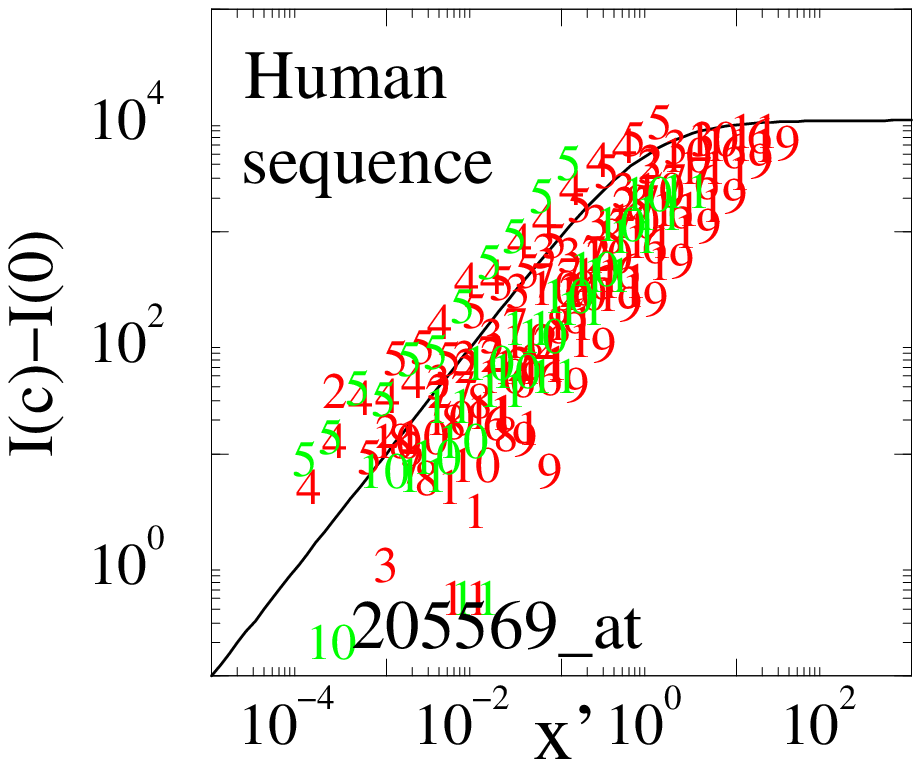}
\includegraphics[width=4.2cm]{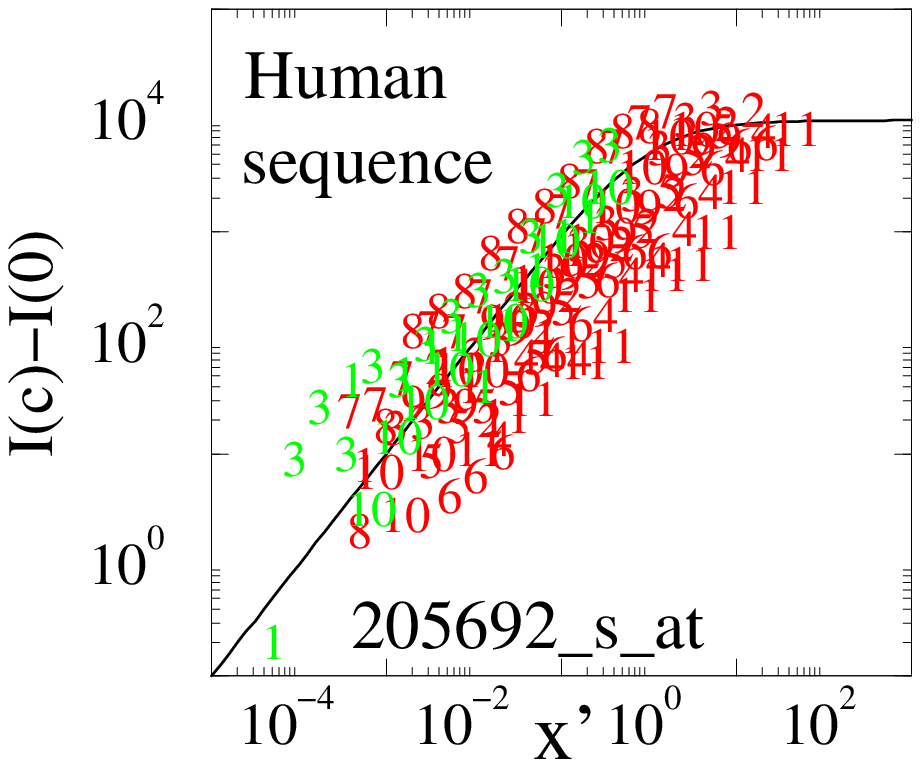}
\includegraphics[width=4.2cm]{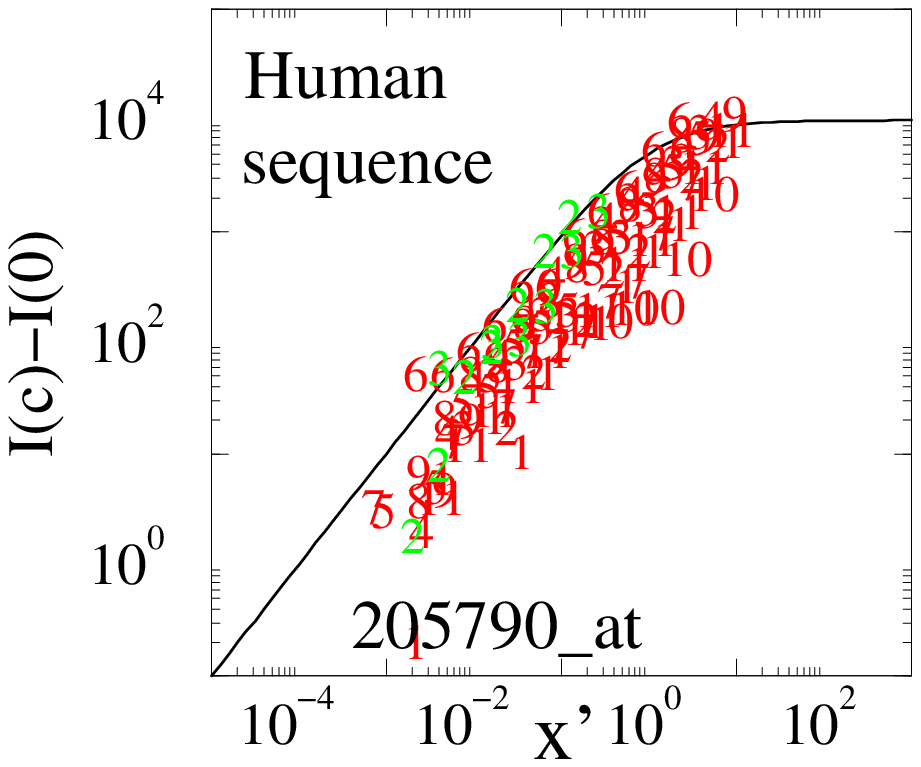}

\includegraphics[width=4.2cm]{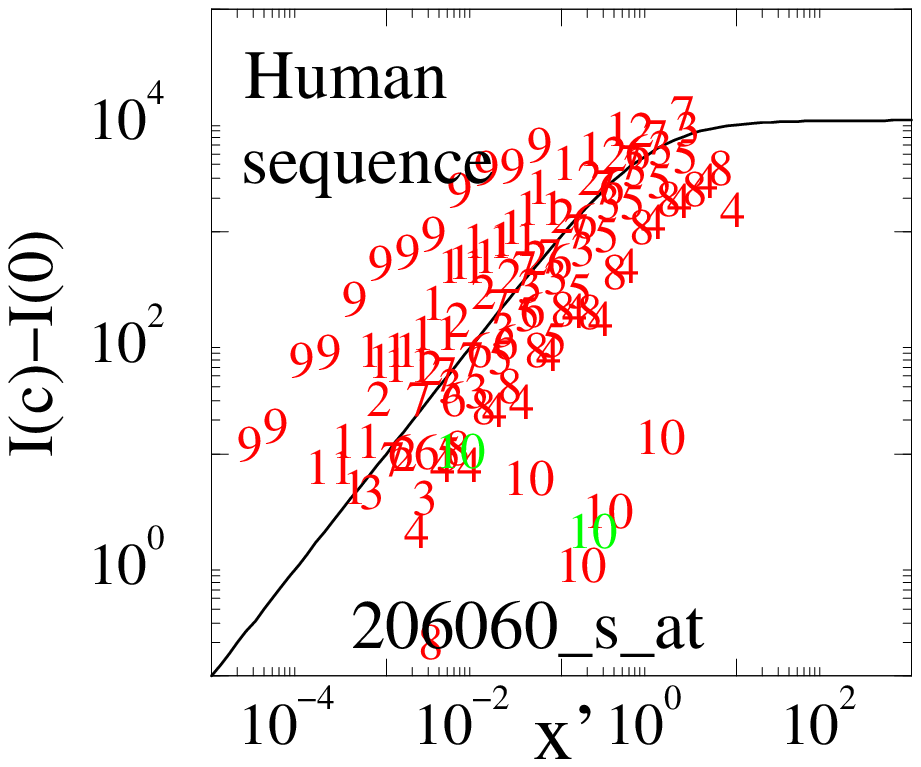}
\includegraphics[width=4.2cm]{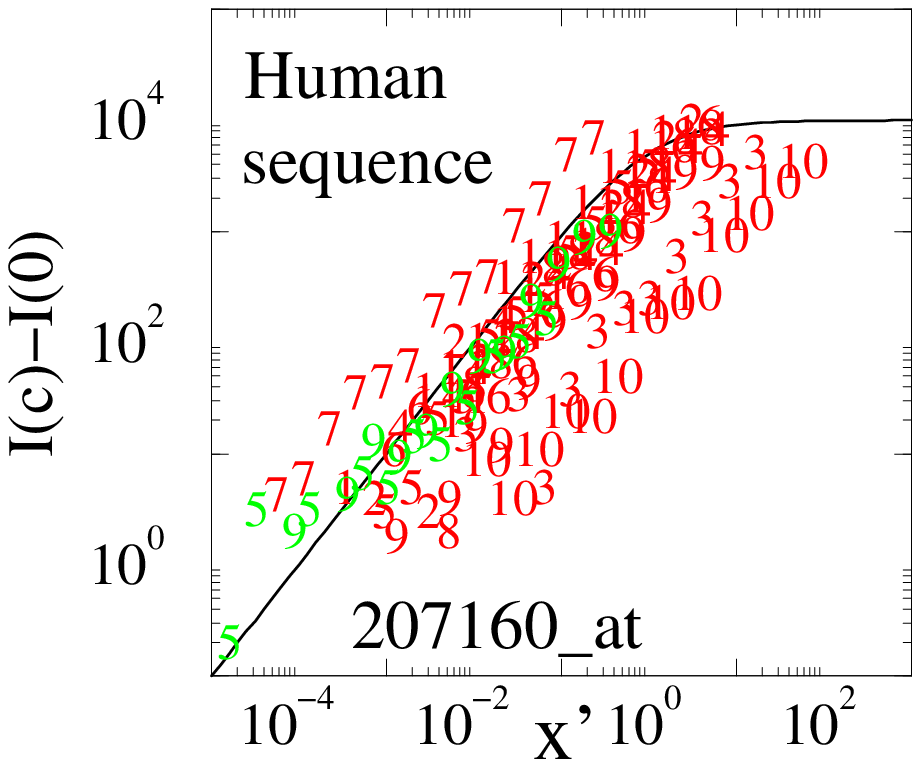}
\includegraphics[width=4.2cm]{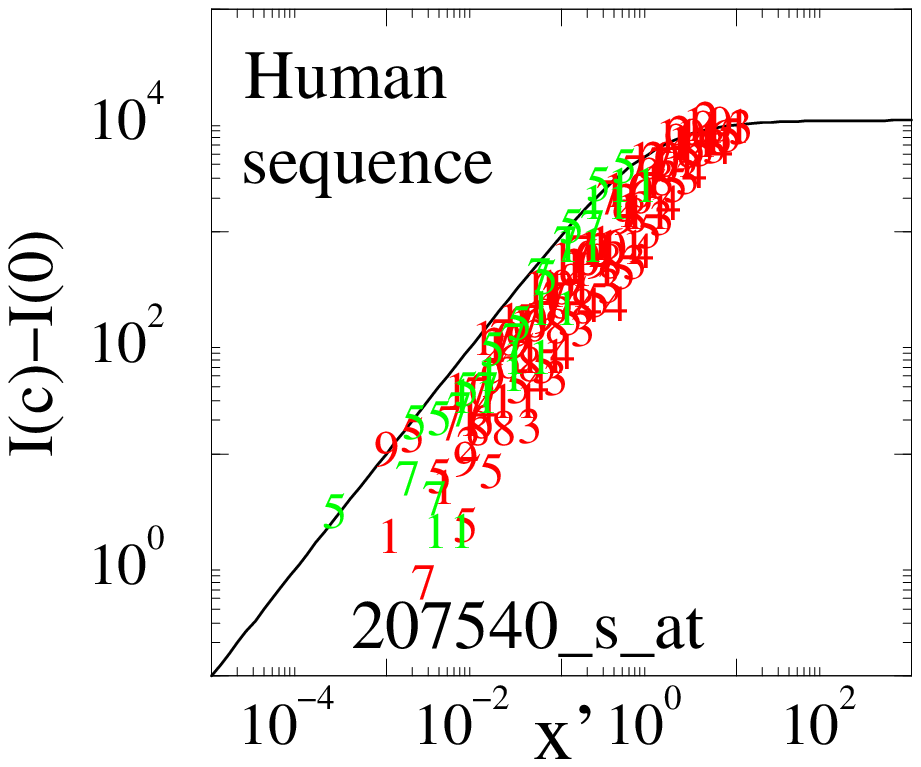}
\includegraphics[width=4.2cm]{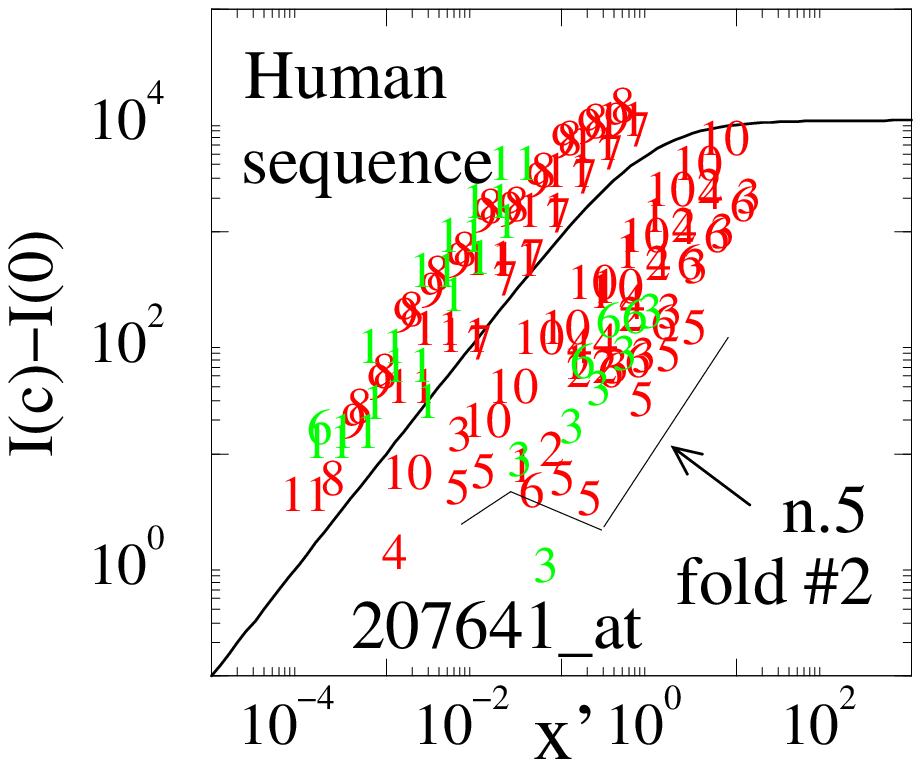}

\includegraphics[width=4.2cm]{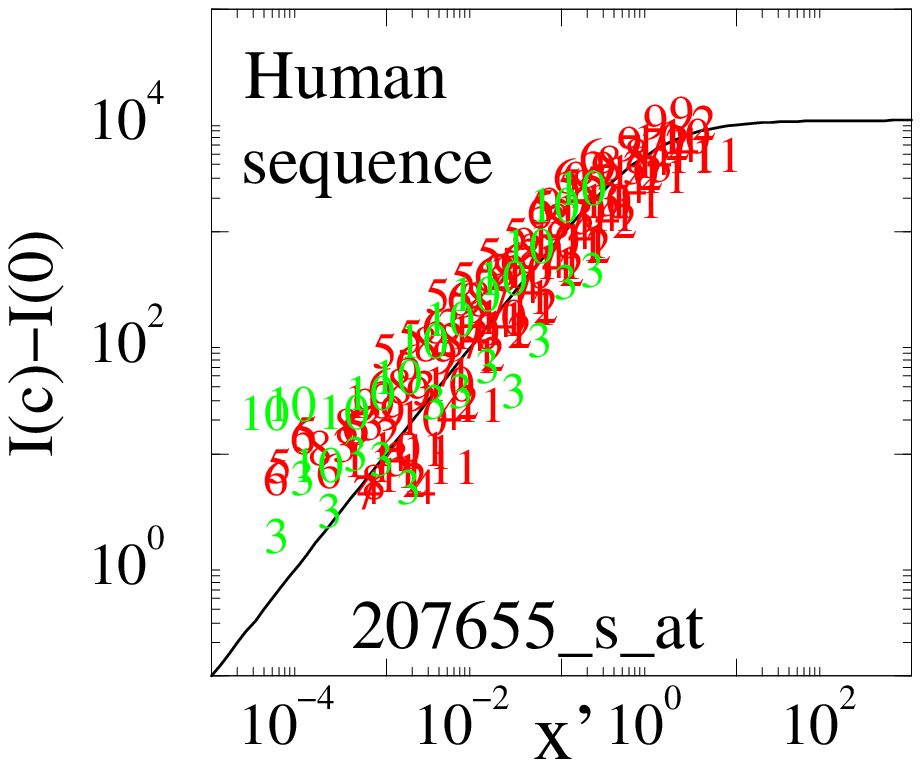}
\includegraphics[width=4.2cm]{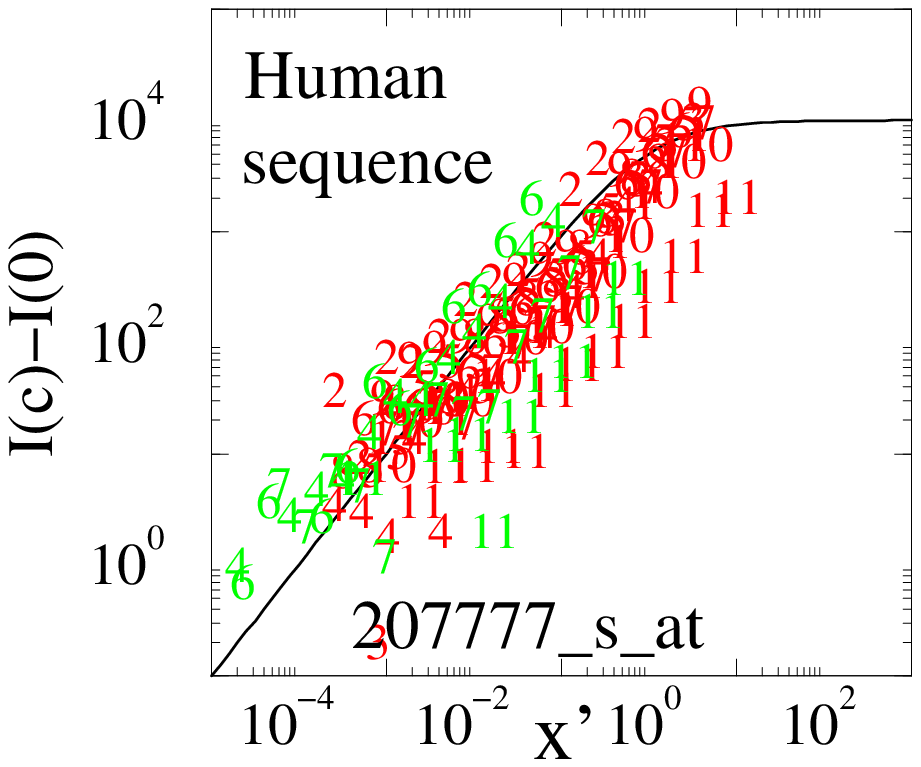}
\includegraphics[width=4.2cm]{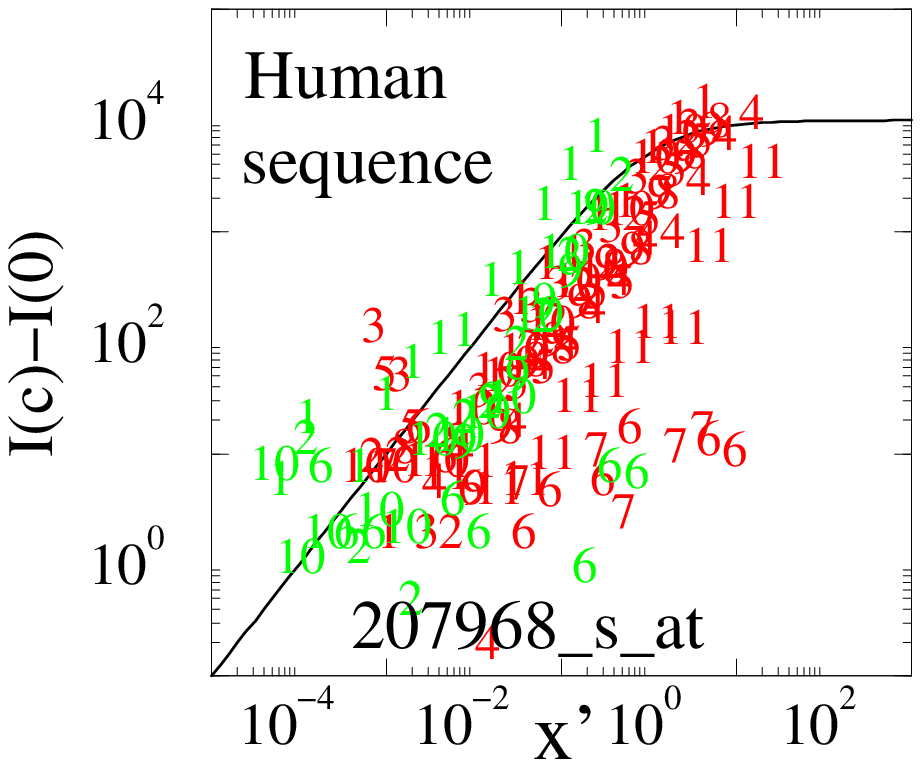}
\includegraphics[width=4.2cm]{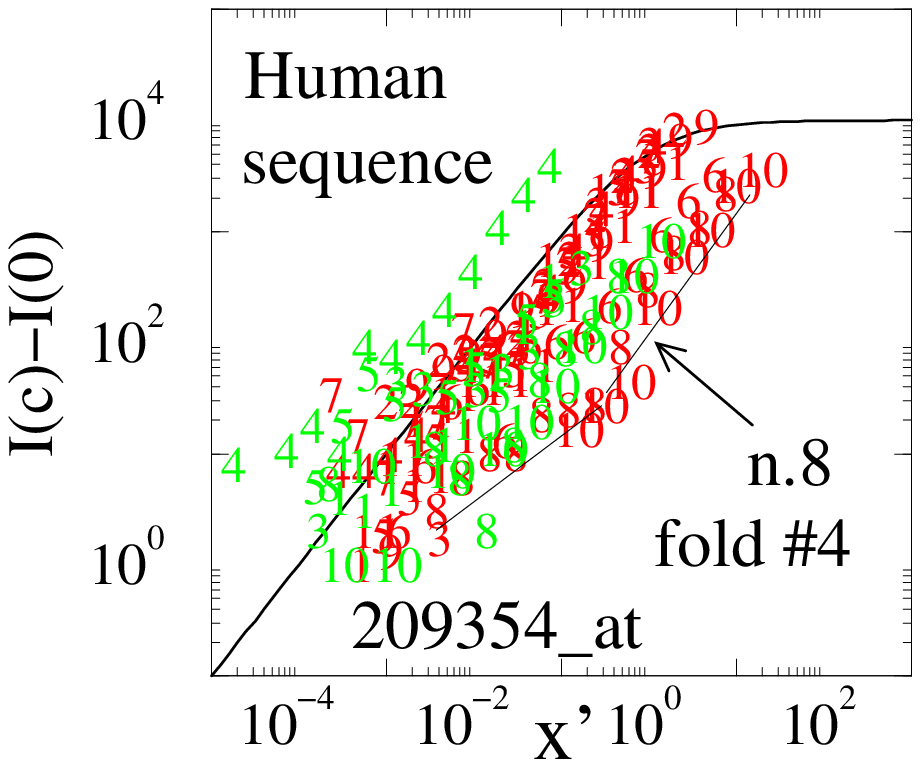}

\includegraphics[width=4.2cm]{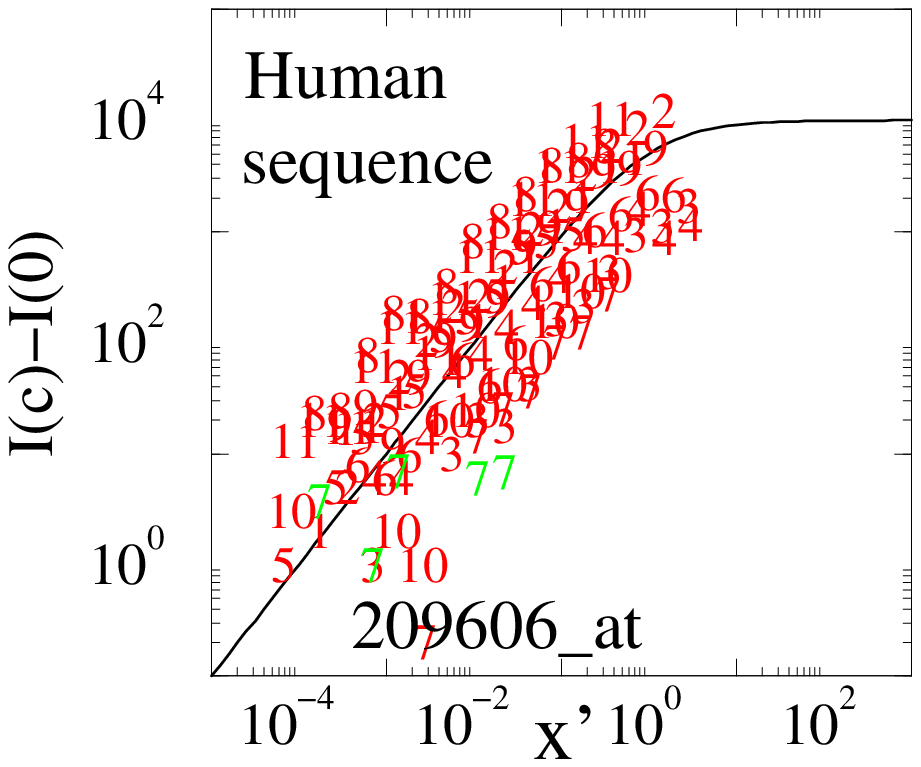}
\includegraphics[width=4.2cm]{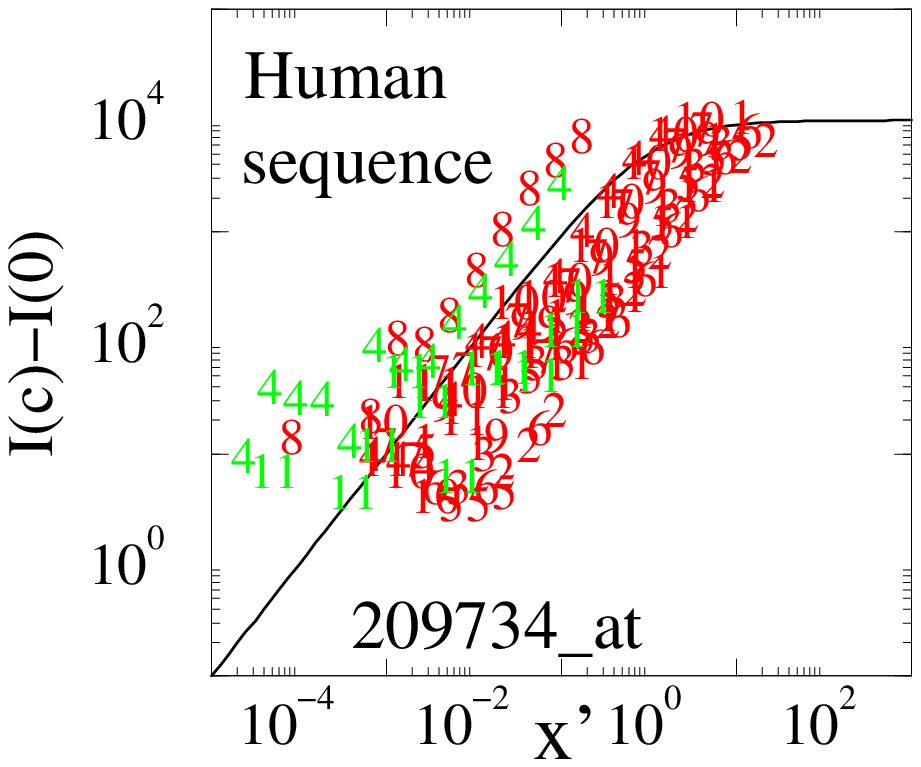}
\includegraphics[width=4.2cm]{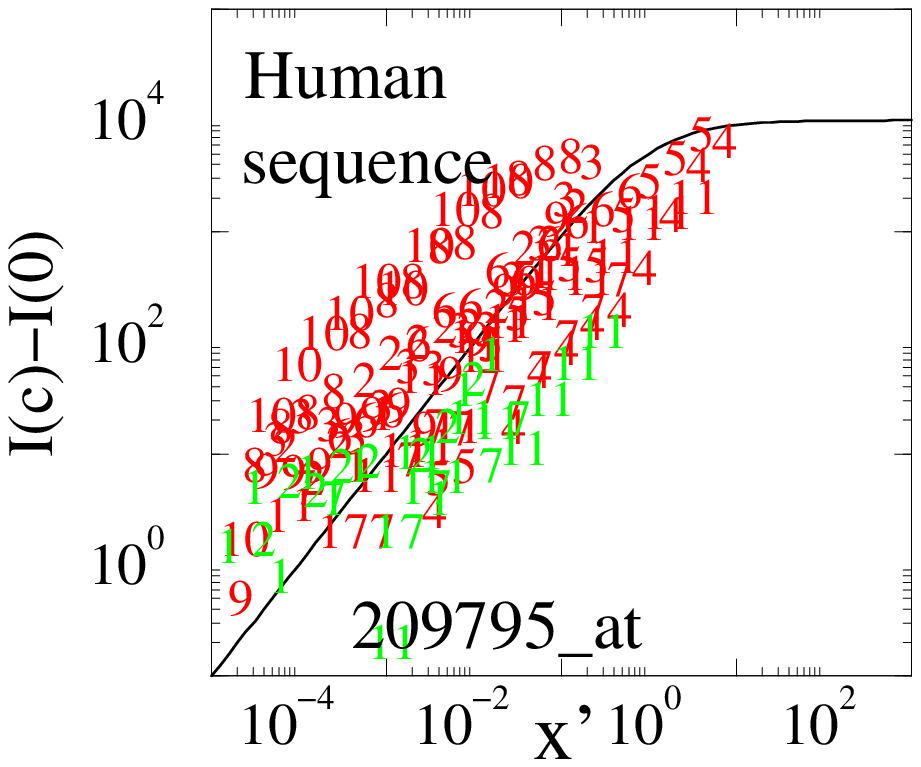}
\includegraphics[width=4.2cm]{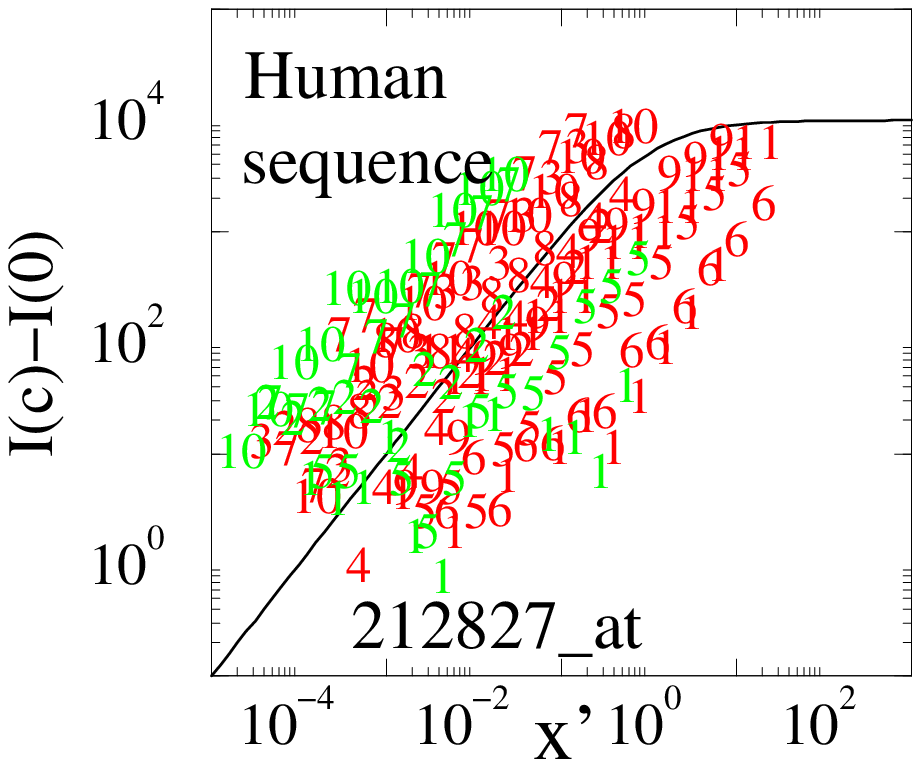}
\caption{Collapse plots for Human sequences of the HGU133 spike-in set 
(part 2). The probes which are complementary to targets which the largest
folding free energies are emphasized (see Table \ref{table_fold}). They
correspond to probes 207641\_at5 and 209354\_at8.
}
\label{collapse_H2}
\end{figure*}

In Figs. \ref{collapse_A}, \ref{collapse_H1} and \ref{collapse_H2}
we show the collapse plots for all the 42 genes of the spike-in data
set HGU133. Each plot contains about 200 points, which all tend to
cluster (in some cases much better than others) along the Langmuir
curve $Ax'/(1+x')$. All the 13 concentrations, which range from $0.125$
pM to $512$ pM in the spike-in experiment, are shown. The intensities
measured at $c=0$ are taken as estimates of the background level $I_0$
in Eq.(\ref{rescaled}).  In the collapse plots only the MM sequences
for which a $\Delta G$ could be estimated are shown, as the mismatch
free energies in RNA/DNA duplexes are known only for a limited set of
mismatches \cite{sugi00} (we could associate a free energy to about 30\%
of mismatches, as discussed in Ref. \cite{carl06}).

The HGU133 spike-in set contains 4 bacterial sequences and 8
artificial sequences (Fig. \ref{collapse_A}) and 30 human sequences
(Fig. \ref{collapse_H1} and \ref{collapse_H2}).  A perfect agreement
with the Langmuir theory would imply that the data all align along the
curve given by Eq. (\ref{rescaled}) and shown as a solid line in the Figs.
\ref{collapse_A}, \ref{collapse_H1} and \ref{collapse_H2}. In general the
agreement is best for the artificial sequences. Occasionally, also some
human sequences collapse well into a single curve in good agreement with
the Langmuir model, but in general their behavior is worse than artificial
ones.  In order to measure the data dispersion we introduce the variable:
\be
w = \log \left( \frac{I}{I_{\rm th}} \right),
\label{define_w}
\ee
where $I$ is the measured intensity and $I_{\rm th}$ the theoretical
value as predicted from the Langmuir isotherm (Eq.~(\ref{rescaled}))
for the $x'$ corresponding to the measured $I$. For the definition of $w$
in Eq. (\ref{define_w}) we have kept only the values of $I$ in the
range $100 < I < 10000$.  We determine its average $\langle w \rangle$
and standard deviation $\sigma_w$. If the data are well-centered around
the expected behavior one has $\langle w \rangle =0$, while $\sigma_w$
is a measure of the spread in the data.

The values of $\langle w \rangle$ and $\sigma_w$ for the
bacterial, artificial and human sequences are given in the tables
\ref{table_bacterial} and \ref{table_human}, respectively.  We note
that $\sigma_w$ is on average the lowest for the artificial sequences
with typical value $\sigma_w \approx 1$. Only for two human probe sets
(205790\_at and 207540\_s\_at with $\sigma_w \approx 0.7$) the collapse
is better than that of the artificial sequences. For three human probe
sets (204205\_at, 207641\_at and 212827\_at) the collapse is very poor
as indicated by a $\sigma_w > 2$. The collapses in the four bacterial
sequences have somewhat higher dispersion compared to human sequences.

A very interesting feature of the whole analysis is that the quality
of collapses is much better for artificial sequences than for any
other sequence. Artificial sequences have been chosen by Affymetrix
to be as different as possible from any human RNA so to minimize the
effects of cross-hybridization. Their preparation, as labeling and
target fragmentation are concerned, is the same as for all other spikes
\cite{private_affy}. As in all collapses the same set of parameters is
used, the high $\sigma_w$ for some probe sets is very likely an indication
that the selected probes are not yet optimal.  Possible deviations from
the theory are due to cross-hybridization.

%%%%%%%%%%%%%%%%%%%%%%%%%%%%%%%%% FIG_01 %%%%%%%%%%%%%%%%%%%%%%%%%%%%%%%%%%%
%%%%%%%%%%%%%%%%%%%%%%%%%%%%%%%%% FIG_01 %%%%%%%%%%%%%%%%%%%%%%%%%%%%%%%%%%%

\section{Determination of the expression level}

\begin{figure}[t]
\includegraphics[width=8.5cm]{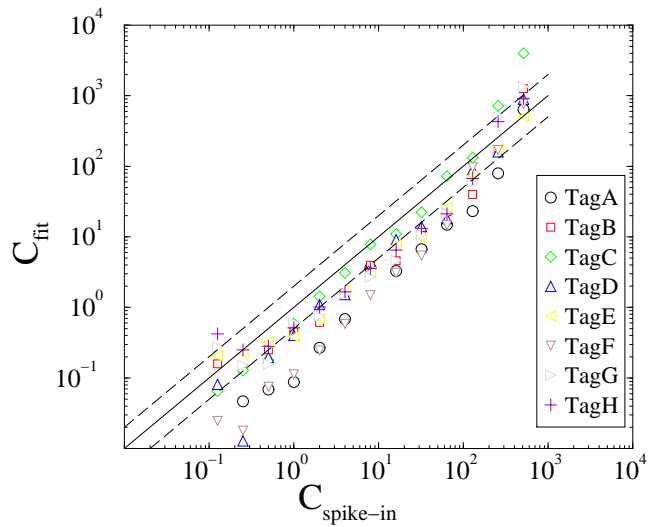}
\caption{Plot of the fitted target concentration as a function of the
spike-in concentration for the artificial sequences. The solid line correspond
to the diagonal $y=x$, while the two dotted lines are $y=x/2$ and $y=2x$
and are drawn as guides to the eye. We note a systematic shift of the 
estimated absolute concentration compared to the spike-in one, although
the fold-variations of the concentrations are correctly estimated
as the majority of the data follow lines parallel to the diagonal in
the plot.}
\label{CvsCspike}
\end{figure}

The model defined by Eqs. (\ref{fluorescence}) and (\ref{alpha}), once
all parameters have been fixed, can be used to fit the concentration
$c$ starting from the measured intensities. The target concentration
in solution is a measurement of the gene expression level and it
is the quantity one wants to compute from the raw microarray data.
As the concentrations in the spike-in experiments are known, we can
compare the known values with the fitted ones. Figure \ref{CvsCspike}
shows a plot of fitted concentration vs. spike-in concentration for the
artificial sequences. We limit ourselves here to show the data for these
sequences, but the trend is quite general and valid for other
genes as well.  The solid line in Fig. \ref{CvsCspike} corresponds to
a line $y=x$, which means perfect agreement between spike-in and fitted
values. The two other lines correspond to $y=2x$ and $y = x/2$, drawn as
a guide to the eye.

As shown in Fig. \ref{CvsCspike}, most of the data fall in the range between
the two lines, except for the spikes TagA and TagF which give a much
lower fitted concentration. All the points follow approximately straight
lines with slope 1, except for the highest spike-in concentrations,
corresponding to $256$ and $512$ pM. This is due to the fact that at
high concentrations many probes are very close to saturation.

We note also that the fitted concentrations are all systematically
lower than the spike-in values, as most of the concentrations fall
in the interval $[c_{\rm spike-in}/2,c_{\rm spike-in}]$. This is a
consequence of our choice to use the fitting parameters which we took from
a previous study \cite{carl06} of spike-in experiments on the HGU95. We
have chosen not to refit these parameters here again for HGU133,
to illustrate their universal validity. The slight underestimation of
the absolute concentration is not a problem, since in gene expression
measurements one is only interested in fold-variations of expression
levels between different experimental conditions. The fact that the data
of Fig. \ref{CvsCspike} follow lines with a slope of approximately one
guarantees that the fold-change in concentration in different experiments
is correctly estimated.

%%%%%%%%%%%%%%%%%%%%%%%%%%%%%%%%% FIG_01 %%%%%%%%%%%%%%%%%%%%%%%%%%%%%%%%%%%
%%%%%%%%%%%%%%%%%%%%%%%%%%%%%%%%% FIG_01 %%%%%%%%%%%%%%%%%%%%%%%%%%%%%%%%%%%

%%%%%%%%%%%%%%%%%%%%%%%%%%%%%%%%%%%%%%%%%%%%%%%%%%%%%%%%%%%%%%%%%%%%%%%%%%%%%%%%
\begin{table}[t]
\caption{Minimal folding free energies for the targets (assumed to be 25-mers)
complementary to the probes forming the spike-in HGU133 data set. These
free energies are calculated with the program RNAfold.}
\begin{ruledtabular}
\begin{tabular}{ccc}
Probe set & Probe number & -$\Delta G_{\rm fold}$(kcal/mol) \\
\hline
204513\_s\_at	& 4	& 8.70 \\
207641\_at	& 5	& 8.16 \\
204430\_s\_at	& 10	& 7.79 \\
209354\_at	& 8	& 7.67 \\
207540\_s\_at	& 10	& 7.45 \\
AFFX-r2-TagA\_at& 1	& 6.52 \\
205398\_s\_at	& 1	& 6.43 \\
AFFX-PheX-3\_at	& 10	& 6.18 \\
204836\_at	& 10	& 6.17 \\
203508\_at	& 2	& 6.10 \\
206060\_s\_at 	& 3	& 6.05 \\
\end{tabular}
\end{ruledtabular}
\label{table_fold}
\end{table}
%%%%%%%%%%%%%%%%%%%%%%%%%%%%%%%%%%%%%%%%%%%%%%%%%%%%%%%%%%%%%%%%%%%%%%%%%%%%%%%%

\section{One cause of outliers: Target secondary structures}

It is well-known that single stranded nucleic acids, particularly RNA,
tend to form stable folded conformations by binding of complementary
bases. Currently, algorithms that calculate RNA secondary structures
are to be trusted for sufficiently short molecules, say less than 50
nucleotides, which is the situation of Affymetrix microarrays, where RNA
targets are fragmented before hybridization. The average target length
is $50$, but probably only shorter fragment contribute to hybridization.

We used the Vienna package \cite{hofa03} for the calculation of folded
RNA structures that may form in solution and impede hybridization.
We considered first 25-mer targets in solution exactly complementary to
the probes of the HGU133 spike-in data set.  Table \ref{table_fold} shows
a list of probes in this set, whose complementary target has the lowest
folding free energy, i.e. that of the most stable conformation, calculated
at the experimental temperature of $45^\circ$ C. Given a folding free energy
$\Delta G_{\rm fold}$, one can use the two state model approximation to
find $p_{\rm fold}$ the probability that the sequence is folded into the
most stable conformation: 
\be 
p_{\rm fold} = \frac{e^{-\Delta G_{\rm fold}/RT}}{1+e^{-\Delta G_{\rm fold}/RT}} 
\label{p_fold} 
\ee 
where we use $T=45^\circ$ C.  According to this expression for a folding
free energy $\Delta G_{\rm fold}= -8$~kcal/mol one finds $1 - p_{\rm
fold} \approx 4\cdot 10^{-6}$ and $\Delta G_{\rm fold}= -6$~kcal/mol $1
- p_{\rm fold} \approx 10^{-4}$. Therefore the large majority of the
targets complementary to the probes listed in Table \ref{table_fold}
are folded and not expected to participate to hybridization.

\begin{figure}[t]
\includegraphics[height=3.6cm]{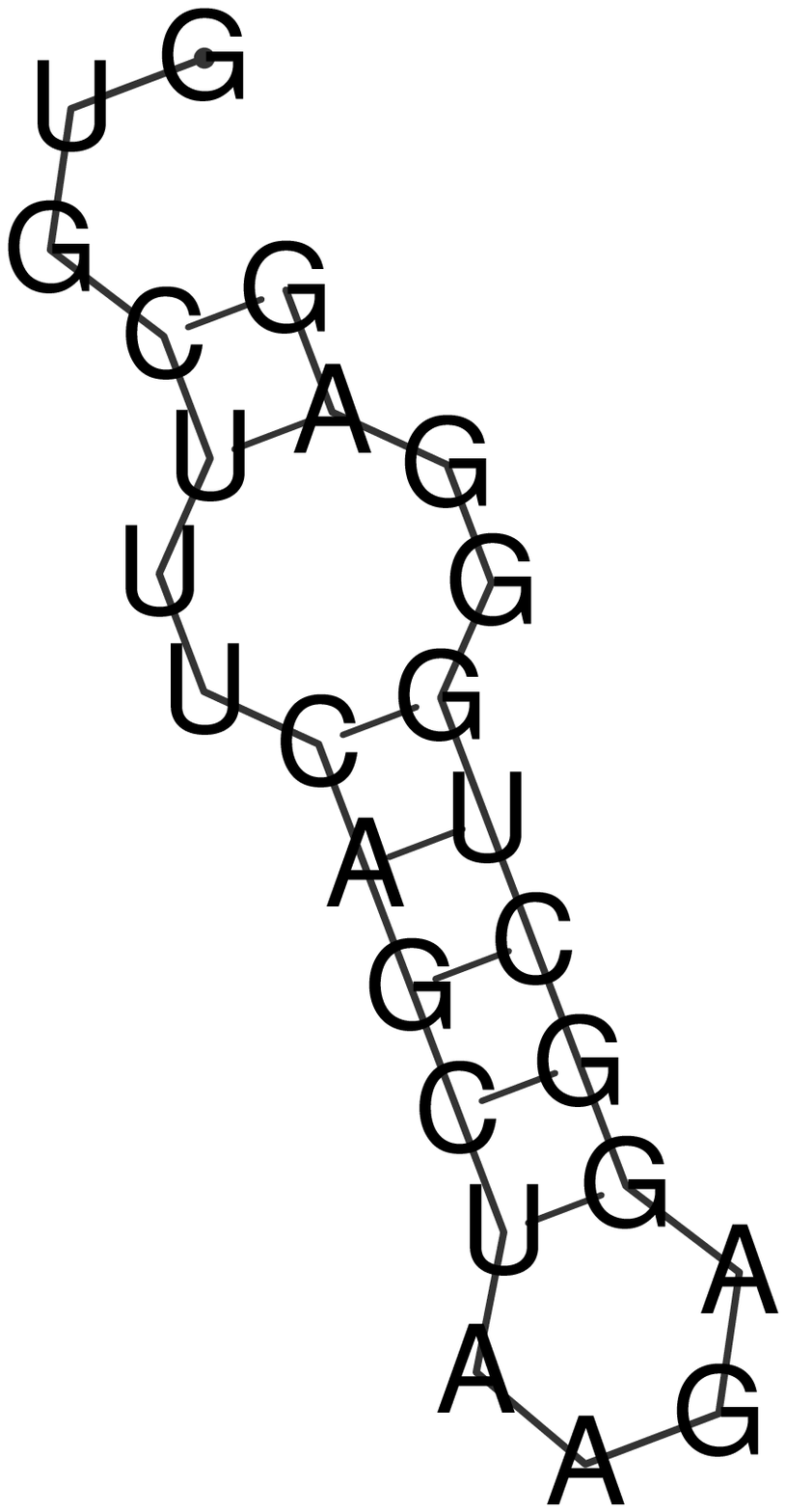}
\includegraphics[height=3.3cm]{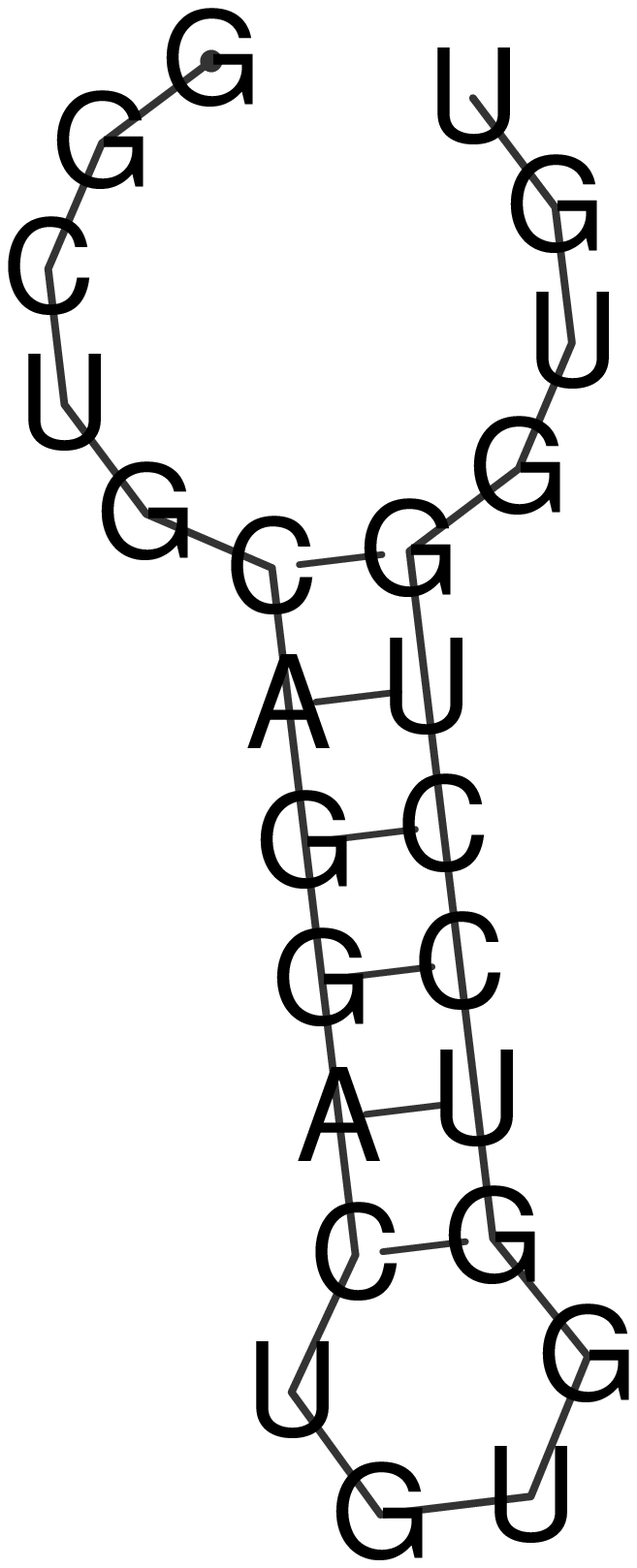}
\includegraphics[height=2.8cm]{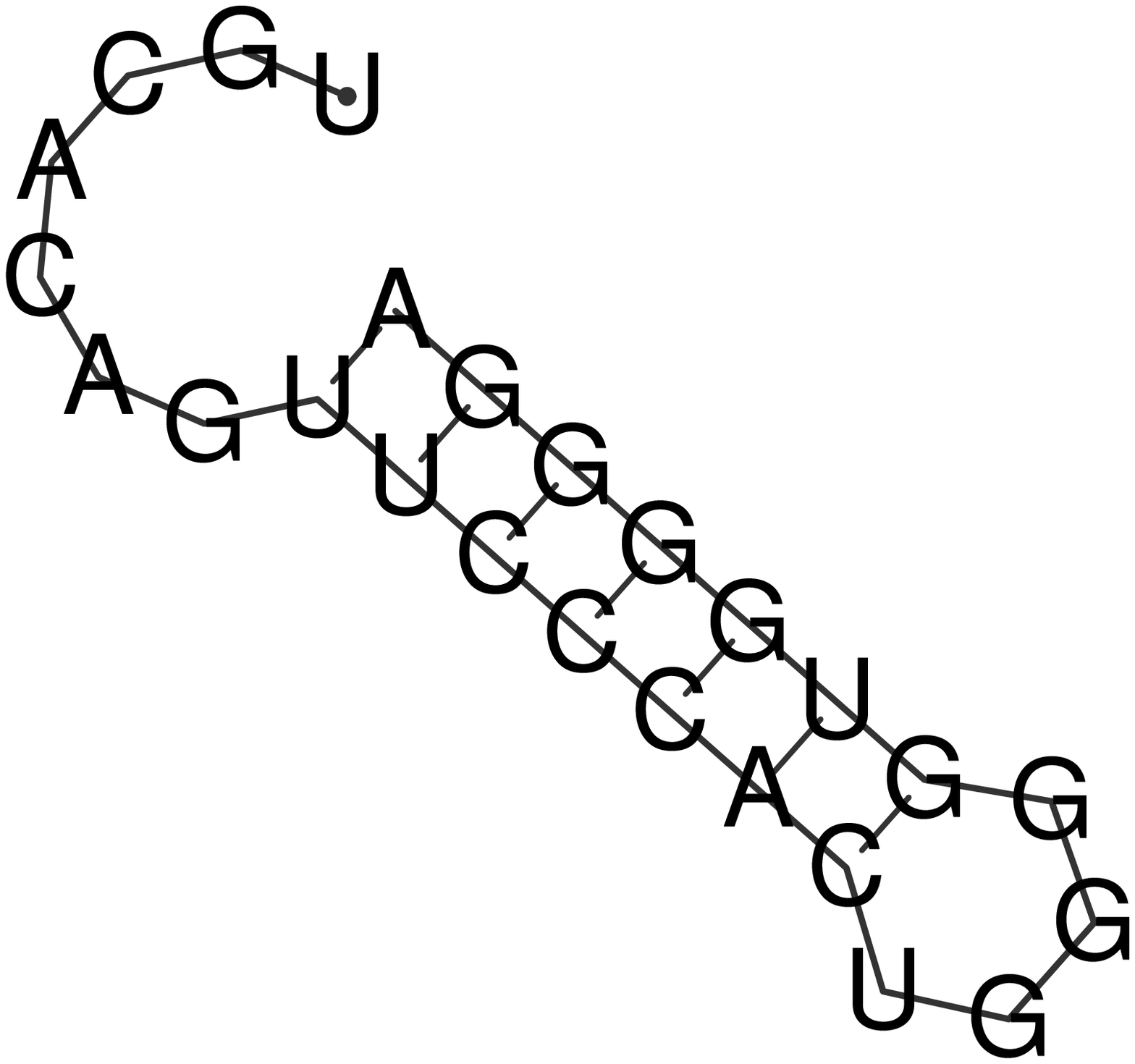}
\includegraphics[height=3.4cm]{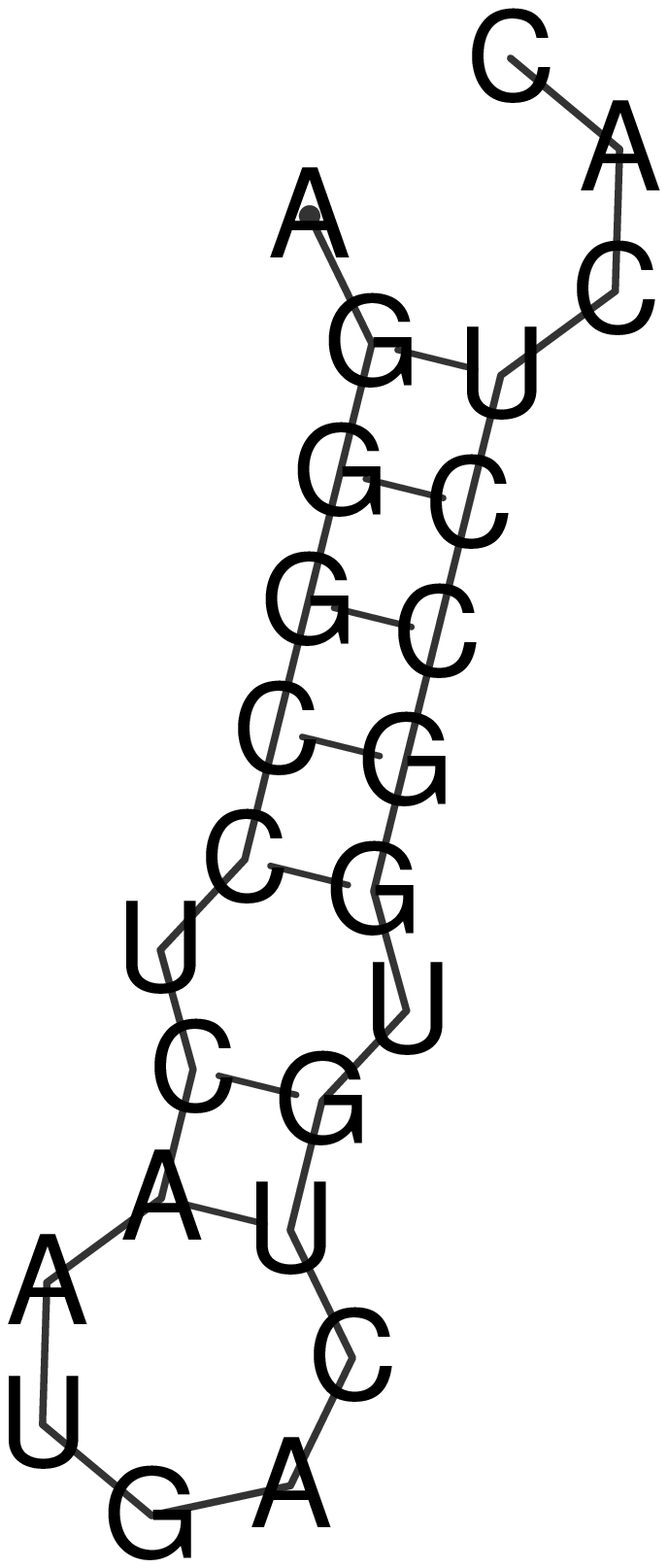}
\caption{Folding configurations for the four targets with the lowest
free energy. From left to right: 204513\_s\_at4, 207641\_at5,
204430\_s\_at10 and 209354\_at8.}
\label{FIG0r}
\end{figure}

Figure \ref{FIG0r} shows the folding configurations for the four
targets with the lowest free energy of Table \ref{table_fold}. As shown
in Figs. \ref{collapse_H1} and \ref{collapse_H2} the corresponding
probes have a signal which is few order of magnitude lower than that
expected from the Langmuir model, although not as low as derived from
Eq. (\ref{p_fold}), using the $\Delta G_{\rm fold}$ listed in Table
\ref{table_fold}.  For instance, from the measured signals we find
an intensity lower by a factor $10^3$ for the probe 204513\_s\_at4,
instead of a factor $10^6$ as deduced from Eq.~(\ref{p_fold}). This
difference could have several origins. First, the hybridization in
solution described by the term $\alpha$ in Eq.~(\ref{alpha}) may
already take into account some secondary structure formation. Second,
the RNA in solution is present with sequences of all lengths. The free
energies listed in Table \ref{table_fold} refer to 25-mers, so shorter
sequences will have lower folding probability than that deduced from
Eq.~(\ref{p_fold}) on the basis of the free energies of 25-mers. Third,
even if some secondary structure is present, hybridization with the
surface-bound probes is still possible if the folded configuration has
some dangling ends from which binding can initiate.

We have analyzed the folding free energies of 25-mers complementary to
all the probes in the HGU spike-in set. We found that about $50\%$ of
the targets have folding free energy lower than $1$ kcal/mol, so that
secondary structure formation can be safely neglected. About $10\%$
of the targets have a folding free energy higher than $4$ kcal/mol, so
that for this fraction the secondary structure formation may interfere
with the target-probe hybridization.

Summarizing, the correct estimate of the folding probability involves
a complex calculation over fragments of all lengths, possibly including
sequences neighboring the 25-mer part complementary to the probe. However
the folding is expected to have a relevant effect for at most $10\%$ of
the probes. A possible way out is that of excluding from the analysis
of the gene expression levels those probes whose 25-mers folding free
energy is above a certain threshold.

%%%%%%%%%%%%%%%%%%%%%%%%%%%%%%%%% FIG_01 %%%%%%%%%%%%%%%%%%%%%%%%%%%%%%%%%%%
%%%%%%%%%%%%%%%%%%%%%%%%%%%%%%%%% FIG_01 %%%%%%%%%%%%%%%%%%%%%%%%%%%%%%%%%%%

\section{Conclusion}

In this paper we have extended a previous study \cite{carl06} of
Affymetrix spike-in experiments on the chip HGU95, to a novel HGU133
chipset. We used the model introduced in Ref. \cite{carl06} which
takes into account both target-probe and target-target hybridization
in solution. The hybridization free energies are calculated from the
nearest-neighbor model \cite{bloo00} using the experimental parameters
for RNA/DNA \cite{sugi95_sh,sugi00} and RNA/RNA \cite{xia98_sh}.
There are four global fitting parameters in the model that we
took from Ref. \cite{carl06}. We found that these parameters fit well
also the current data on the HGU133 chipset, apart for a systematic
small shift of all the estimates of the absolute target concentrations.

There are several features that make the spike-in data of the more recent
HGU133 chip interesting. First of all the spike-in set contains a larger
number of sequences compared to the HGU95 experiments (42 instead of 14)
and the chip has been entirely redesigned. Secondly, the spike-in
sequences contain some of artificial origin, designed to avoid any
cross hybridization with human RNAs, but prepared and labeled exactly
as all other spikes.  We find that these artificial sequences fit best
the hybridization model, as they show the best collapses when the data
are rescaled and plotted as function of an appropriate thermodynamic
variable. The good agreement suggests indeed that the simple model
describes rather well the hybridization in Affymetrix arrays and that
the deviations observed for some human sequences are probably related to
the non-optimal design of the sequences for a given probe.

When compared to the human sequences of the HGU95 spike-in experiments
analyzed in Ref. \cite{carl06}, we find that the artificial spikes of
the HGU133 set show definitely better collapses.  However, when comparing
the human sequences of the HGU133 with those in the HGU95 experiment we
find on average a better collapse for the latter. Only few probes out
of the 32 human spikes of the HGU133 experiment have a better collapse
than those of the HGU95.

Interestingly, the physics-based modeling developed here allows to assign
to each probe set a quality score based on the level of agreement on the
Langmuir model. This information may be used to reconsider and eventually 
redesign the probe sets of low quality.

Finally, we have discussed the physical basis of hybridization in solution
and of RNA secondary structure formation. The latter effect, according
to the statistics over the spike-in probes, will be relevant for about
10\% of the probes only. The sequences with the highest folding probability
correspond to probes whose measured fluorescent intensities is well-below
that predicted from the Langmuir model.

According to our current understanding of the system (see also
Refs. \cite{carl06,heim06}), the hybridization in solution of
partially complementary RNA molecules has a strong influence.  One of
the reasons for that is that RNA/RNA interaction parameters are, at
given temperature and salt concentration, stronger than the DNA/DNA or
RNA/DNA parameters. The simple approximation given in Eq.(\ref{alpha})
captures the major features of the hybridization in solution. However,
an improvement over this approach, as discussed above, remains an open
challenge.

We acknowledge financial support from the Van Gogh Programme d'Actions
Int\'egr\'ees (PAI) 08505PB of the French Ministry of Foreign Affairs
and the NWO grant 62403735.

% \bibliography{/home/enrico/TEX/biblio.bib}

\end{document}